\documentclass[pra,twocolumn]{revtex4}
\usepackage{amsfonts,amsmath,amssymb,float}
\usepackage{bm,dsfont}
\usepackage{color}
\usepackage{bbm}
\usepackage[dvips]{epsfig}
\usepackage{amsmath,amssymb,lscape,float}
\usepackage{hyperref}
\usepackage{listings}
\usepackage{lipsum}
\usepackage{diagbox}

\usepackage{graphicx}% Include figure files
\usepackage{epstopdf}
\usepackage{dcolumn}% Align table columns on decimal point
\usepackage{bm}% bold math
\usepackage{amsfonts,amsmath,amssymb%,float
}
\usepackage{todonotes}
\usepackage{braket}
\usepackage{verbatim}
\usepackage{siunitx}
\usepackage{xcolor}

\renewcommand{\vec}[1]{\mathbf{#1}}
\renewcommand{\vec}[1]{\ensuremath{\boldsymbol #1 }}

\newcommand{\mrm}{\mathrm}
\newcommand{\mn}{\mathnormal}

\DeclareSIUnit\betaz{\beta_\mrm{z}}
\DeclareSIUnit\Er{\mrm{E_R}}
\DeclareSIUnit\SystemEnergy{\hbar\mrm{\omega_z}}
\DeclareSIUnit\SystemTime{\mrm{\omega_z}^{-1}}

\begin{document}
	\title{Single-atom transport in optical conveyor belts:
	Enhanced shortcuts-to-adiabaticity approach}

	\author{Sascha H. Hauck, Gernot Alber, and Vladimir M. Stojanovi\'c}
	\affiliation{Institut f\"{u}r Angewandte Physik, Technical
		University of Darmstadt, D-64289 Darmstadt, Germany}
	\date{\today}
	\begin{abstract}
	 Fast and nearly lossless atomic transport, enabled by moving the confining trap,
	 is a prerequisite for many quantum-technology applications. While theoretical
	 studies of this problem have heretofore focussed almost exclusively on simplified scenarios
	 (one-dimensional systems, purely harmonic confining potentials, etc.), we investigate it
	 here in the experimentally relevant setting of a moving optical lattice ({\em optical
	 conveyor belt}). We model single-atom transport in this system by taking fully
	 into account its three-dimensional, anharmonic confining potential. We do so using the
	 established method of shortcuts to adiabaticity (STA), i.e. an inverse-engineering
	 approach based on Lewis-Riesenfeld invariants, as well as its recently proposed modification 
         known as {\em enhanced} STA (eSTA). By combining well-controlled, advanced analytical 
         techniques and the numerical propagation of a time-dependent Schr\"{o}dinger equation 
         using the Fourier split operator method, we evaluate atom-transport fidelities within 
         both approaches. Being obtained for realistic choices of system parameters, our results 
         are relevant for future experiments with optical conveyor belts. Moreover, they reveal 
         that in the system at hand the eSTA method outperforms its STA counterpart for all but 
         the lowest optical-lattice depths.
	\end{abstract}

	\maketitle
	\section{Introduction}
	Efficient transport of cold neutral atoms~\cite{Couvert+:08,Murphy+:09,Masuda+Nakamura:10,Chen+:10,
	Torrontegui+:11,Chen+:11,Ness+:18,Amri+:19,Hickman+Saffman:20,Ding+:20,Lam+:21} -- either in the form of
	condensates or individually -- is of utmost importance for a variety of emerging quantum-technology
	applications~\cite{Navez+:16,Rodriguez-Prieto+:20} as well as for quantum-state engineering~\cite{Ebert+:14,
	StojanovicPRA:21,Haase+:21}.
	Such transport, often referred to as ``shuttling''~\cite{Qi+:21} and expected to be fast and nearly
	lossless, entails moving the confining magnetic-~\cite{Nakagawa+:05,Purdy+:10,Pandey+:19} or optical
	trap~\cite{Kuhr+:03,Sauer+:04}. In particular, moving optical traps come in two varieties -- moving
	optical lattices~\cite{Hickman+Saffman:20,Lam+:21} and tweezers~\cite{Barredo+:18,Brown+:19,Browaeys+Lahaye:20}.
	For single-atom transport it is typically required that the final atomic state be as close as possible
	to the initial one -- up to an irrelevant global phase factor -- in the rest frame of the moving trap
	(the high-fidelity condition). This is equivalent to demanding complete absence (or, at least, minimization)
	of vibrational excitations at the end of transport, but does not rule out the existence of transient
	excitations at intermediate times~\cite{Torrontegui+:11}.

	The lack of requirement for adiabaticity throughout atom-transport processes motivates the use of control
	protocols known as shortcuts to adiabaticity (STA)~\cite{STA_RMP:19} for their modelling. Generally speaking,
	the latter lead to the same final states as slow, adiabatic changes of the control parameters of a system,
	but typically require significantly shorter times to reach that state. This makes the system much less
	prone to the debilitating effects of noise and decoherence. Importantly, adiabatic processes are those
	for which slow changes of control parameters leave some dynamical properties of the system invariant.
	As a consequence, arguably the most useful ones among STA methods are inverse-engineering techniques
	based on the concept of Lewis-Riesenfeld invariants~\cite{Lewis+Riesenfeld:69}.

	While STA protocols have already found applications in a variety of quantum systems~\cite{STA_RMP:19}, their
	analytical modification -- inspired by optimal-control techniques~\cite{Werschnik+Gross:07} -- has quite recently
	been proposed and termed {\em enhanced shortcuts to adiabaticity} (eSTA)~\cite{Whitty++}. The main motivation
	behind eSTA is to design efficient control protocols for systems to which STA protocols are not directly
	applicable. The principal idea of eSTA is to first approximate the full Hamiltonian of such a system by
	a simpler one for which an STA protocol can be found. Assuming that this STA protocol for the simplified
	Hamiltonian is close to being optimal even when applied to the full system Hamiltonian, the actual optimal
	eSTA protocol is obtained through a gradient expansion in the space of control parameters.
	In principle, the heuristic character of eSTA does not guarantee its superiority over STA
	and, indeed, the criteria as to when this scheme can be expected to work efficiently are still unknown~\cite{Whitty++}.
	Yet, eSTA has already been shown to outperform STA in certain problems of moderate complexity~\cite{Whitty++},
	which motivates its use in more complex problems.

        Theoretical studies of coherent single-atom transport have heretofore relied on simplified
        scenarios, typically assumming a one-dimensional geometry (i.e. motion only along the direction
        of transport)~\cite{Torrontegui+:11,ZhangEtAl:15} or taking the confining potential to be purely
        harmonic~\cite{Murphy+:09,Chen+:11}. However, in realistic systems such idealizations often do
        not apply, either because there is a significant coupling between the longitudinal and transverse
        degrees of freedom or because the trapping potential is strongly anharmonic. An important example
        of such systems is furnished by {\em optical conveyor belts} (OCBs)~\cite{Schrader+:01,Kuhr+:03},
        moving optical lattices enabled by two counterpropagating Gaussian laser beams with equal intensities,
        which are slightly mutually detuned. Those systems constitute powerful tools for the precise positioning
        of atoms~\cite{Schrader+:01,Sauer+:04,Fortier+:07}, with the added capabilities to enable high-speed
        transport over macroscopic distances and quickly sort atoms into ordered arrays~\cite{Dotsenko+:05,
        Miroshnychenko+:06,Langbecker+:18}. In the context of single-atom transport, OCBs have been investigated
        quite recently~\cite{Hickman+Saffman:20,Lam+:21}.

	In this paper, we address single-atom transport in an OCB using both STA and eSTA methods. We model this
	system by taking fully into account its underlying three-dimensional (3D), anharmonic confining potential.
	Using an existing inverse-engineering single-atom transport theory~\cite{Torrontegui+:11}, we first obtain
	an STA solution for the trajectory of a moving trap. We then obtain -- by combining the obtained STA solution
	with advanced analytical techniques -- an eSTA solution of the same problem. Finally, based on the
	designed trap trajectories we evaluate the resulting single-atom dynamics through the numerical propagation of
	a time-dependent Schr\"{o}dinger equation. We quantify these dynamics by computing atom-transport fidelities
	for a broad range of lattice depths within both STA and eSTA frameworks.

	Given that they correspond to realistic choices of the relevant experimental parameters (beam waists, lattice
	depths, transport distances, etc.), our obtained results are of utmost interest for future experiments with OCBs.
	Furthermore, these results show that the eSTA method yields faster single-atom transport than STA for all but
	the lowest optical-lattice depths.

The remainder of this paper is organized as follows. In Sec.~\ref{system} we introduce the system
under consideration and its characteristic length-, time-, and energy scales. Section~\ref{MovingTrap}
discusses the design of trajectories of the moving trap. This is first done using the STA method,
i.e. a Lewis-Riesenfeld invariant (Sec.~\ref{MovingTrap_STA}), and then using the eSTA method based
on the STA solution for a harmonically-approximated OCB potential (Sec.~\ref{MovingTrap_eSTA}).
In Sec.~\ref{AtomDynamics} we briefly describe our methodology for computing the resulting single-atom
dynamics. We first review the general aspects of the Fourier split operator method
(Sec.~\ref{FSOMbasics}), followed by specific details of our implementation thereof (Sec.~\ref{ComovingFrameFSOM}).
Our findings are presented and discussed in Sec.~\ref{ResultsDiscussion}, starting with the
atom-transport fidelities obtained for a broad range of system parameters using the STA and eSTA methods
(Sec.~\ref{Fidelity_STA_eSTA}), and followed by a comparison of the latter results with alternative approaches
(Sec.~\ref{SineAndTriangleComp}). We conclude, with a short summary of the paper and some general remarks,
in Sec.~\ref{SummaryConclusions}. Some involved mathematical derivations are relegated to Appendices
\ref{derivationG_n} and \ref{derivationK_n}, while Appendix \ref{tableAppendix} summarizes certain
intermediate calculation results.
\section{System and its Hamiltonian} \label{system}
We consider an atom of mass $m$ in an OCB, whose optical axis is in the $z$ direction.
In what follows, we will be concerned with the problem of transporting an atom to a distant
location -- the distance being at least an order of magnitude larger than the size of the atomic
wave packet -- along this same (longitudinal) direction. This mimics the physical situation
encountered in typical experimental setups~\cite{Hickman+Saffman:20,Lam+:21}. The initial- and
target atomic states are assumed to be the ground states of the OCB potential centered at
two different locations.

The relevant single-atom Hamiltonian reads
\begin{equation}\label{SingleAtomHamiltonian}
H_{\textrm{OCB}}=-\frac{\hbar^2\nabla^2}{2m} + U_{\textrm{F}}(x,y,z)\:,
\end{equation}
where $U_{\textrm{F}}(x,y,z)$ is the full 3D potential of an OCB:~\cite{Hickman+Saffman:20}
\begin{eqnarray}\label{eqPotentialOpticalConveyorBelt}
U_{\textrm{F}}(x,y,z) &=& U_{f,0}(z) \cos^2(kz) \nonumber \\
&\times& \exp\left(-2 \left[\frac{x^2}{w_x(z)^2} +
\frac{y^2}{w_y(z)^2} \right] \right)\:.
\end{eqnarray}
Here $k=2\pi/\lambda$ is the wave number of the dipole-trap laser with
wavelength $\lambda$. The lattice depth $U_{f,0}(z)$ is given by
\begin{equation}\label{exprU_0}
U_{f,0}(z)=C\:\frac{P_0}{w_x(z)w_y(z)}\:,
\end{equation}
where $w_{x}(z)$, $w_{y}(z)$ are the two transverse beam waists, which
depend on the longitudinal position $z$:
\begin{equation}\label{zDependentWaists}
w_{x/y}(z)=w_{x/y,0}\:\sqrt{\displaystyle
1+\left(\frac{z}{Z_{R,x/y}}\right)^2} \:,
\end{equation}
with $Z_{R,x}$ and $Z_{R,y}$ being the respective Rayleigh lengths. In Eq.~\eqref{exprU_0}
$P_0$ stands for the output laser power, while the constant $C=\hbar\Gamma^2/(2\Delta\:I_0)$
characterizes the concrete experimental setup, with the saturation intensity $I_0$, the decay
rate $\Gamma$, and the detuning $\Delta=\omega-\omega_0$ between the laser frequency $\omega$
and the frequency $\omega_0$ of the relevant atomic transition (e.g. $\omega_0 = 2\pi\times\:384.23$
\:THz for the Rubidium $D_2$-line~\cite{Steck:21}).

Finding a harmonic approximation $V(x,y,z)$ of the full OCB potential $U_{\textrm{F}}$ in Eq.~\eqref{eqPotentialOpticalConveyorBelt}
is of crucial interest for our further considerations. This simplified potential can readily be found
by applying a harmonic approximation to various terms in $U_{\textrm{F}}$. To this end, we first assume
$z/Z_{R,x} \ll 1$ and $z/Z_{R,y} \ll 1$. We also assume that $x/w_{x,0}\ll 1$, $y/w_{y,0}\ll 1$, and
that $k \, z \ll 1$. Under these assumptions, it is straightforward to find that
\begin{equation}\label{eqSimplePotential}
V(x,y,z)=-U_0 +\frac{m}{2}\left(\omega_x^2 x^2 +
\omega_y^2 y^2 + \omega_z^2 z^2 \right)\:,
\end{equation}
where $U_0 \equiv U_{f,0}(0)$ is the potential depth at the focus of the beam and
the frequencies $\omega_x$, $\omega_y$, and $\omega_z$ are respectively given by
\begin{eqnarray}\label{eqFreqs}
\omega_x^2 &=& \frac{4U_0}{m w_{x,0}^2} \:, \nonumber \\
\omega_y^2 &=& \frac{4U_0}{m w_{y,0}^2} \:, \\
\omega_z^2 &=& \frac{U_0}{m} \left( Z_{R,x}^{-2}
+ Z_{R,y}^{-2} + 2\:k^2 \right) \nonumber \:.
\end{eqnarray}
It is useful to note that in the paraxial approximation $Z_{R,x/y}\gg 1/k$,
which is always valid for OCBs, one has that $\omega_z^2\approx 2 U_0 k^2/m$.
By taking into account Eqs.~\eqref{eqFreqs} and the well-known relation
$Z_{R,x/y}=kw_{x/y,0}^2/2$, one concludes that there are five independent
parameters in the system at hand: the transport distance $d$, the final
time $t_f$, the waists $w_{x/y,0}$, and the potential depth $U_0$.

To facilitate our further discussion, it is prudent to single out the characteristic
time-, length-, and energy scales in the system under consideration.
The time $\tau_{\textrm{z}}=2\pi/\omega_{\textrm{z}}$ corresponding to the harmonic-oscillator frequency
$\omega_{\textrm{z}}$ in the $z$ direction [cf. Eq.~\eqref{eqSimplePotential}] will be used in what
follows as the characteristic timescale. On the other hand, the harmonic-oscillator length $l_{\textrm{z}}
\equiv\sqrt{\hbar/(2m\omega_z)}$ in the $z$ direction will serve as the characteristic lengthscale.
Finally, all energies in the problem will be expressed in units of the recoil energy
$E_{\textrm{R}}\equiv\hbar^2 k^2/(2m)$.

\section{STA and $\textrm{e}$STA trap trajectories} \label{MovingTrap}
Among all STA methods~\cite{STA_RMP:19}, invariant-based inverse engineering
established itself as the method of choice in the context of efficient atom
transport. The basic invariant-based inverse engineering transport theory was
developed in Ref.~\cite{Torrontegui+:11}. The crux of that theory is the use of
quadratic-in-momentum invariants relevant for transport problems, which
were first discussed by Lewis and Riesenfeld~\cite{Lewis+Riesenfeld:69}.
Importantly, it was also demonstrated in Ref.~\cite{Torrontegui+:11} that
the case of a harmonic trapping potentials and that of an arbitrary potential
require different treatments, as the perfect atom transport in the latter
case necessitates -- in principle -- compensating forces in the reference
frame moving with the trap (cf. Sec.~\ref{ComovingFrameFSOM} below).

In the following, we first apply the theory developed in Ref.~\cite{Torrontegui+:11}
to our problem of single-atom transport in OCBs. To be more precise, we determine
the classical path of the potential minima in a moving trap in the problem at hand
(Sec.~\ref{MovingTrap_STA}). We then apply the eSTA scheme, based on the theory recently
laid out in Ref.~\cite{Whitty++}, to address the same problem (Sec.~\ref{MovingTrap_eSTA}).
We do so by making use of a single-atom Hamiltonian with the harmonically approximated
OCB potential $V(x,y,z)$ [cf. Eq.~\eqref{eqSimplePotential}] as the simplified Hamiltonian
of the system for which an STA-based protocol can readily be obtained.

\subsection{Trajectory of the moving trap: STA solution} \label{MovingTrap_STA}
A dynamical invariant of a time-dependent Hamiltonian $H(t)$ is an operator $I(t)$,
which satisfies the equation
\begin{equation}\label{eqLewisRiesenfeldInvariant}
\frac{\partial}{\partial t} I(t) + \left[H(t),I(t)\right] = 0 \:.
\end{equation}
The eigenvalues $\lambda_n$ of the invariant $I(t)$ are constant in time. Assuming that
these eigenvalues are non-degenerate, the corresponding eigenstates $\ket{\Phi_n(t)}$ and
the instantaneous eigenstates $\ket{\Psi_n(t)}$ of the Hamiltonian $H(t)$ (the so-called
transport modes) satisfy the simple relation $\ket{\Psi_n(t)}=e^{i\theta_{\textrm{LR}}(t)}\ket{\Phi_n(t)}$,
where $\theta_{\textrm{LR}}(t)=\hbar^{-1}\int_0^t\bra{\Phi_n(t')}\left[i\hbar\partial_{t'}-H(t')\right]
\ket{\Phi_n(t'} dt'$ is the Lewis-Riesenfeld phase~\cite{STA_RMP:19}. The general solution of the
Schr\"{o}dinger equation for the Hamiltonian $H(t)$ can then be written in the form
\begin{equation}\label{eqGeneralSolutionSchroedinger}
|\Psi(t)\rangle= \sum_{n} C_n\: e^{i\theta_{\textrm{LR}}(t)}|\Phi_n(t)\rangle \:.
\end{equation}
It is worth noting that for very long evolution times ($t\rightarrow\infty$)
Eq.~\eqref{eqLewisRiesenfeldInvariant} results in the adiabatic-following condition
$\left[H(t),I(t)\right] \approx 0$.

In what follows, we apply Lewis-Riesenfeld theory to the approximate OCB Hamiltonian
\begin{equation}\label{simpleOCB_Hamiltonian}
H_0=-\frac{\hbar^2\nabla^2}{2m}+V(x,y,z-q_0(t))\:,
\end{equation}
i.e. a single-atom Hamiltonian with the simplified harmonic potential $V(x,y,z)$
of Eq.~\eqref{eqSimplePotential}. For our transport scheme we make use of the
time-dependent, quadratic-in-momentum invariant~\cite{Torrontegui+:11}
\begin{equation}
I = \frac{1}{2 m}\:\left(p-m\dot{q}_{c,z}\right)^2+
\frac{m}{2}\:\omega_z^2(z-q_{c,z})^2 \:,
\end{equation}
where $q_{c,z}$ is the $z$ component of the classical path for the trapped particle.
Importantly, there are auxiliary equations that must be fulfilled in order to use this
invariant~\cite{Torrontegui+:11}. For simple displacement schemes the auxiliary equation
has the form characteristic of a forced harmonic oscillator. It reads
\begin{equation}\label{eqForcedHarmonicOscillator}
\begin{split}
\ddot{q}_{c,z}(t) + \omega_z^2 \left[ q_{c,z}(t) - q_0(t) \right]=0 \:,
\end{split}
\end{equation}
where $q_0(t)$ is the trajectory of the potential minimum.

In order to fulfill the appropriate boundary conditions for the ``classical'' particle, we are
choosing a polynomial Ansatz of ninth degree, by which the general solution for the path of
the potential minima can be obtained through Eq.~\eqref{eqForcedHarmonicOscillator}.
This results in
\begin{equation}
q_0(t) = d \sum_{n=3}^9 b_n \left( \frac{t}{ t_f } \right)^n
\end{equation}
with the following solution vector for constants $b_n$:
\begin{equation}\label{eqSolutionvectorPotential}
\begin{split}
\vec{X}_{q_0} \, = \,
\begin{pmatrix}
b_3 \\
b_4 \\
b_5 \\
b_6 \\
b_7 \\
b_8 \\
b_9
\end{pmatrix}
=
\begin{pmatrix}
2520(t_f \omega_z)^{-2} \\
-12600(t_f \omega_z)^{-2} \\
22680(t_f \omega_z)^{-2} +126 \\
-17640(t_f \omega_z)^{-2} -420 \\
5040(t_f \omega_z)^{-2} +540 \\
-315 \\
70
\end{pmatrix}\:.
\end{split}
\end{equation}

\begin{figure}[t!]
\includegraphics[clip,width=0.875\linewidth]{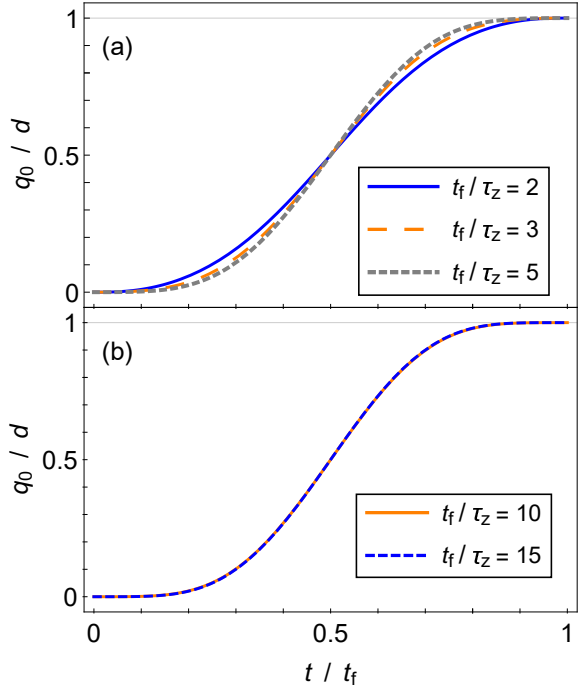}
\caption{\label{fig:STAminimum}(Color online) Path of the potential minimum as a function of time, obtained
using the STA approach, for transport times $t_f$ (a) comparable to, and (b) an order of magnitude longer 
than the internal timescale $\tau_{\textrm{z}}$.}
\end{figure}

The obtained classical path of the potential minimum for different final times $t_f$ is
shown in Fig.~\ref{fig:STAminimum}.

\subsection{Trajectory of the moving trap: eSTA solution} \label{MovingTrap_eSTA}
Generally speaking, the first step in applying the eSTA method to a system with the Hamiltonian
$H_{\textrm S}$ entails obtaining an STA solution for a ``close'' Hamiltonian $H_0$~\cite{Whitty++};
this solution is parameterized by a vector $\vec{\lambda}_0 \in \mathbb{R}^n$. In the present context,
the meaning of ``close'' is that there exists a parameter $\mu_{\textrm S}$ such that $H_{\textrm S}$
can be expressed in the form of a series expansion
\begin{equation} \label{eqCloseHamiltonian}
H_{\textrm S} = \sum_{k=0}^\infty \mu_{\textrm S}^k \, H^{(k)} \:,
\end{equation}
such that $H^{(0)}\equiv H_0$. In the OCB system at hand, the role of $H_{\textrm S}$ is played by the
Hamiltonian $H_{\textrm{OCB}}$ of Eq.~\eqref{SingleAtomHamiltonian}, in which $z$ is replaced by $z-q_0(t)$.
On the other hand, the role of $H_0$ is played by the Hamiltonian of Eq.~\eqref{simpleOCB_Hamiltonian}.

Because we aim to find an optimal solution for the full Hamiltonian $H_{\textrm S}$ based on an available
STA solution for $H_0$, it is prudent to express the general control vector for the full system in the form
$\vec{\lambda}=\vec{\lambda}_0 +\vec{\alpha}$, i.e. as a sum of the STA control vector $\vec{\lambda}_0$ and
an auxiliary control vector $\vec{\alpha}$. The value of $\vec{\alpha}$ that corresponds to the optimal
solution, i.e. the correction vector necessary to obtain the optimal eSTA protocol will be denoted by
$\vec{\epsilon}$ in the following.

A crucial assumption within the eSTA scheme is that the protocol based on the existing STA
solution for $H_0$ is close to being optimal even when applied to the full system Hamiltonian
$H_{\textrm S}$~\cite{Whitty++}. Furthermore, one assumes that the fidelity is quadratic
around its maximal value, resulting in the approximate relation~\cite{Whitty++}
\begin{equation} \label{eqESTAFidelity}
F\left(\mu_{\textrm{S}},\vec{\lambda}_0 + \alpha \frac{\vec{\nabla}F(\mu_{\textrm{S}},\vec{\lambda}_0)}
{\rVert\vec{\nabla}F(\mu_{\textrm{S}},\vec{\lambda}_0)\rVert}\right)\approx 1 - c\left(\alpha-\epsilon
\right)^2 \:,
\end{equation}
where $\epsilon\equiv\rVert\vec{\epsilon}\rVert$, $\alpha\equiv\rVert\vec{\alpha}\rVert$, and $c$
is a positive constant. Using the above assumptions and a Taylor expansion of the left-hand-side of 
Eq.~\eqref{eqESTAFidelity} around $\alpha=\epsilon$, it is straightforward to find that~\cite{Whitty++}
\begin{equation}\label{eqInitialEpsilon}
\vec{\epsilon}\approx\frac{2\left[1- F(\mu_{\textrm{S}},\vec{\lambda}_{\textrm{S}})\right]
\vec{\nabla}F(\mu_{\textrm{S}},\vec{\lambda}_0)}{\rVert\vec{\nabla}F(\mu_{\textrm{S}},
\vec{\lambda}_0)\rVert^2} \:.
\end{equation}
As derived in Ref.~\cite{Whitty++}, the fidelity can be approximated up to second order
in $\mu_{\textrm S}$ as
\begin{equation}\label{eqFidelityGn}
F(\mu_{\textrm{S}},\vec{\lambda}_{\textrm{S}}) \approx
1-\frac{1}{\hbar^2}\sum_{n=1}^\infty |G_n|^2 \:,
\end{equation}
with $G_n$ being an auxiliary (scalar) function, given by
\begin{equation}\label{eqExpressionG}
G_n = \int_0^{t_f} dt \braket{\Psi_n(t)|
\left[H_{\textrm{S}}(\vec{\lambda}_0; t) - H_0
(\vec{\lambda}_0; t)\right]|\Psi_0(t)} \:,
\end{equation}
and $\ket{\Psi_n(t)}$ the transport modes of the idealized Hamiltonian $H_0$ [cf.
Sec.~\ref{MovingTrap_STA}]. An analogous approximate expression, up to second
order in $\mu_{\textrm S}$, for the gradient of the fidelity reads~\cite{Whitty++}
\begin{equation}\label{eqGradientFidelityKn}
\vec{\nabla}F(\mu_{\textrm{S}},\vec{\lambda}_0) \approx
-\frac{2}{\hbar^2} \sum_{n=1}^\infty \textrm{Re}
\left( G_n \, \vec{K}_n^* \right) \:,
\end{equation}
where $\vec{K}_n$ is another auxiliary (vector) function:
\begin{align}\label{eqExpressionK}
\vec{K}_n = \int_0^{t_f} dt \braket{\Psi_n(t)|\nabla_\lambda H_{\textrm{S}}
(\vec{\lambda};t)\big|_{\vec{\lambda}=\vec{\lambda}_0}|\Psi_0(t)} \:.
\end{align}

The optimal correction vector $\vec{\epsilon}$ can be recast in terms of the auxiliary
functions $G_n$ and $\vec{K}_n$ as
\begin{equation}\label{eqEpsilonDefinition}
\vec{\epsilon}= -\frac{\left(\sum_{n=1}^\mrm{N} |G_n|^2 \right) \sum_{n=1}^{N}
\textrm{Re} \left( G_n^\ast \vec{K}_n \right)}{\left\rVert \sum_{n=1}^{N} \textrm{Re}
\left( G_n^\ast \vec{K}_n \right) \right\rVert^2} \:,
\end{equation}
where $N$ is the cut-off parameter. This vector can be computed numerically
once the expressions for $G_n$ and $\vec{K}_n$ are obtained by evaluating the integrals in
Eqs.~\eqref{eqExpressionG} and \eqref{eqExpressionK}, respectively. In the atom-transport
problem at hand, where the states $\ket{\Psi_n(t)}$ represent the transport modes of the 3D
harmonic-oscillator Hamiltonian in Eq.~\eqref{simpleOCB_Hamiltonian}, this entails highly nontrivial
derivations based on various properties of Hermite polynomials (for details, see 
Appendices~\ref{derivationG_n} and \ref{derivationK_n}).

For displacement schemes the control vector $\vec{\lambda}$ has to fulfill the
conditions $q_0(\vec{\lambda}; \, j t_f/7) = \lambda_j$ for $j=1,\ldots, 6$.
Now the optimized path can be expressed through the path of the simplified problem
\begin{equation}\label{eqControlVectorRedefined}
q_0(\vec{\lambda}; \, t) = q_0(\vec{\lambda_0}; \,t) + f(\vec{\alpha}; \, t) \:,
\end{equation}
with $q(\vec{\lambda}; \, jt_f/7) = \lambda_{0,j} + \alpha_j$ for $j=1,\ldots, 6$.
The auxiliary function $f(\vec{\alpha};\:t)$ has to obey the following boundary conditions:
\begin{equation}
\begin{split}
& f(\vec{\alpha};0) = f(\vec{\alpha};t_f) = 0\:, \\
& f(\vec{\alpha}; jt_f/7 ) = \alpha_j \quad (\:j=1,\ldots,6\:) \:,\\
& \frac{ \mrm{d^{(n)}} }{\mrm{d}t^{(n)}}f(\vec{\alpha};t')
|_{t'=\lbrace 0,t_f \rbrace} = 0 \quad (\:n=1,\ldots,4\:)\:.
\end{split}
\end{equation}
The latter conditions are chosen such that $f(\vec{\alpha};\:t)$ can be controlled
through $\vec{\alpha}$ and also obeys the conditions of continuity. Therefore, we
choose the following polynomial Ansatz of eleventh degree:
\begin{equation}\label{eqSolutionvectorF}
f(\vec{\alpha};t) = \sum_{n=0}^{11}\sum_{k=1}^6 \tilde{a}_{n,k}\alpha_k
\left( \frac{t}{t_f} \right)^n.
\end{equation}
The specific values for the coefficients $\tilde{a}_{n,k}$ in the last equation are
given in Table I in Appendix~\ref{tableAppendix}.

For the optimal eSTA solution, we set the auxiliary control vector $\vec{\alpha}$ equal to
the optimal correction vector $\vec{\epsilon}$. The latter can be calculated using the general
expression in Eq.~\eqref{eqEpsilonDefinition}. Because in our 3D problem the transport modes
can be enumerated using three 1D quantum numbers $\{n_x,n_y,n_z\}$, we can rewrite the sum in
Eq.~\eqref{eqEpsilonDefinition} in terms of the main quantum number $n$ and $\{n_x,n_y,n_z\}$.
For the cut-off parameter we take the value $N=2$, even though our numerical evaluation shows
that already taking $N=1$ yields essentially the same result.

\begin{figure}[t!]
\includegraphics[clip,width=0.875\linewidth]{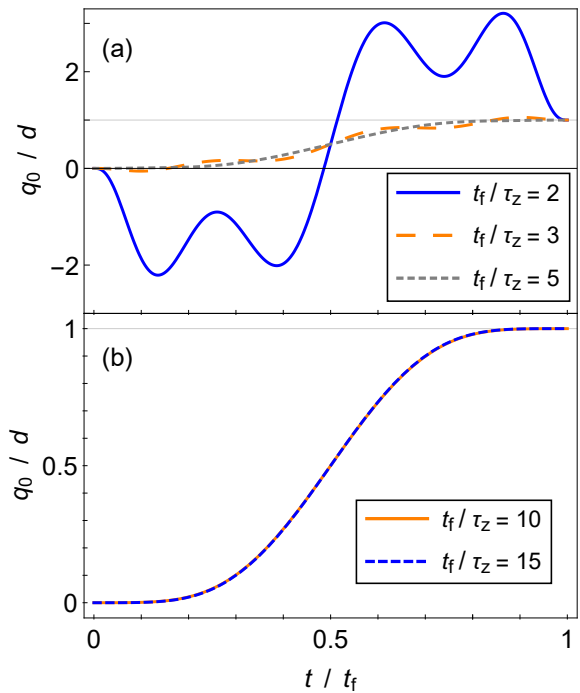}
\caption{\label{fig:eSTAminimum}(Color online) Path of the potential minimum as a function of time,
obtained using the eSTA approach, for transport times $t_f$ (a) comparable to, and (b) an order of
magnitude longer than the internal timescale $\tau_{\textrm{z}}$. The parameters chosen are
the following: $U_0 = 60\:E_{\textrm{R}}$, $d = 85\:l_z$, and $w_{0,x} = w_{0,y} = 4.2\times 10^6\:l_z$.}
\end{figure}

The classical path of the potential minimum, obtained using the eSTA approach, is depicted in
Fig.~\ref{fig:eSTAminimum}. What can be inferred by comparing this path to the one obtained
using the STA approach (Fig.~\ref{fig:STAminimum}) is that their shapes differ significantly
only for short transport times.

\section{Single-atom dynamics: the Fourier split operator method} \label{AtomDynamics}
Having described the design of trap trajectories within both STA and eSTA schemes in Sec.~\ref{MovingTrap},
in the following we briefly present our chosen approach for evaluating the resulting single-atom dynamics
using the Fourier split operator method (FSOM). We start with a brief reminder of the basics of the FSOM
(Sec.~\ref{FSOMbasics}), followed by the specific details of our own implementation thereof to the
single-atom transport problem (Sec.~\ref{ComovingFrameFSOM}).
\subsection{Basics of the FSOM} \label{FSOMbasics}
The FSOM is customarily used for solving Cauchy-type initial-value problems of the type
\begin{equation}\label{CauchyIVP}
\frac{\partial}{\partial t}f(\mathbf{x},t)=\hat{A}(t)f(\mathbf{x},t) \:,
\end{equation}
with some (possibly time-dependent) operator $\hat{A}(t)$ and the initial condition
$f(\mathbf{x},t)=f_0(\mathbf{x})$. The method is typically used in cases where the operator
$\hat{A}(t)$ can be written as a sum $\hat{A}(t)=\hat{A}_1(t)+\hat{A}_2(t)$ of two operators,
such that $\hat{A}_1(t)$ can easily be diagonalized in real space, while $\hat{A}_2(t)$ is
straightforward to diagonalize in Fourier space. In particular, the time-dependent Schr\"{o}dinger
equation (hereafter abbreviated as TDSE) fulfills this requirement, hence the
use of the FSOM for solving this type of equations~\cite{Feit+:82,Bandrauk+Shen:93}.
This requirement allows one to approximate the time-evolution operator of the
system by a product of operators that are diagonal either in real- or in Fourier
space, the central idea of the FSOM.

In the problem at hand, we make use of the FSOM to determine the final atomic state after
displacement by a certain distance. The relevant TDSE corresponds to the Hamiltonian of the type
$H(\vec{r},t)=-\hbar^2\nabla^2/(2m)+W(\vec{r},t)$, where the potential $W$ pertains to a moving
trap and is, consequently, time-dependent (for details, see Sec.~\ref{ComovingFrameFSOM} below).
As a result, the exact time-evolution operator of the system is given by the most general
expression that involves a time-ordered product.

By expanding the time-evolution operator $U(t+\delta t,t)$ of the system to third order
in $\delta t$, we obtain:
\begin{equation}\label{eqFSOMProblem}
U(t+\delta t,t)=\exp\left[-\frac{i}{\hbar}\int_{t}^{t+\delta t}
H(\vec{r},t)dt\right]+O[(\delta t)^3]\:.
\end{equation}
By making use of the Baker-Campbell-Hausdorff formula~\cite{GilmoreBOOK:12}, the last
equation gives an explicit second-order accurate time-stepping scheme for the propagation
of the wave-function $\Psi(\vec{r},t)$~\cite{Pechukas+Light:66}:
\begin{eqnarray}\label{eqFSOMBCH}
\Psi(\vec{r},t&+&\delta t) = \exp\left[-\frac{i}{\hbar}\:W(\vec{r},t)
\frac{\delta t}{2}\right]\exp\left(i\frac{\hbar\nabla^2}{2m}\:\delta t\right) \nonumber \\
&\times& \exp\left[-\frac{i}{\hbar}\:W(\vec{r},t)\frac{\delta t}{2}\right]
\Psi(\vec{r},t)+ O[(\delta t)^3]  \:.
\end{eqnarray}
The last equation allows one to treat the different exponential terms independently, resulting
in the possibility of Fourier-transforming the kinetic term to momentum space. As a result, the
complexity of applying an operator on the wave-function $\Psi$ reduces to that of multiplying
$\Psi$ by a complex number. Importantly, one can recast the right-hand-side of Eq.~\eqref{eqFSOMBCH}
using the identity
\begin{eqnarray}\label{eqFourierRepresentation}
&&\exp\left(i\frac{\hbar\nabla^2}{2m}\:\delta t\right)\exp\left[-\frac{i}{\hbar}\:W(\vec{r},t)
\frac{\delta t}{2}\right]\Psi(\vec{r},t)=  \\
&&F^{-1}\left[\exp\left(-i\frac{\hbar k^2}{2m}\delta t\right)F\left
[\exp\left[-\frac{i}{\hbar}W(\vec{r},t)\frac{\delta t}{2}\right]
\Psi(\vec{r},t)\right]\right] \nonumber \:,
\end{eqnarray}
where $F$ is the Fourier transform and $F^{-1}$ its inverse.

A general solution at time $t'=t+N_t\delta t$ is obtained numerically by applying the single-step
propagation of Eq.~\eqref{eqFSOMBCH} consecutively $N_t$ times to our initial wave-function
$\Psi(\vec{r},t)$. In an actual numerical implementation of the FSOM, this last wave-function is
discretized on a rectangular regular lattice of $N_s$ points and the continuous Fourier transform is
approximated by a discrete one. The computational complexity of propagating the function $\Psi(\vec{r},t)$
is dominated by the transformation into Fourier space and back into real space [cf. Eq.~\eqref{eqFourierRepresentation}]. 
If these transformations are carried out using the fast Fourier transform (FFT) algorithm~\cite{NRcBook}, 
an elementary step in the FSOM requires $O(N_s\log_{2}N_s)$ operations.

Apart from using the FSOM for computing single-atom dynamics, we also utilize this method to
find the ground state of our OCB trapping potential~\cite{Feit+:82}. Let $\phi(\vec{r})=
\sum_{j=0}^N c_j\Psi_j(\vec{r})$ be an arbitrary trial state with a nonzero overlap with the
sought-after ground state $\Psi_0(\vec{r})$. Assuming that $\phi(\vec{r})$ is the initial ($t=0$)
state in a dynamical evolution of the system, its counterpart at a later time $t$ is given by
\begin{equation}
\phi(\vec{r},t)=\sum_{j=0}^N \exp\left(-i E_j t\right/\hbar) c_j\Psi_j(\vec{r}) \:.
\end{equation}
By switching from real to imaginary time, i.e. performing a Wick rotation into the complex plane,
this last state can be recast as a sum of exponentially-decaying contributions of different eigenstates
$\Psi_j(\vec{r})$, with the decay rates given by the corresponding eigenvalues $E_j$. Because the relative
contribution of the excited states decays faster than that of the ground state, these contributions become
negligible for long enough evolutions. This enables one to extract the desired ground-state energy $E_0$
and the corresponding wave function $\Psi_0(\vec{r})$.
\subsection{TDSE in the comoving frame} \label{ComovingFrameFSOM}
Due to time restrictions and storage capabilities, we are restricting ourselves to the displacement
of one single trap minimum. Furthermore, we are switching from the lab frame to the {\em comoving}
frame, i.e. the frame moving along with the trap. This change is accounted for by applying the unitary
displacement operator~\cite{GottfriedYanBOOK:03}
\begin{equation}\label{eqUnitaryOperatorTrapFrame}
\mathcal{U} = \mathrm{e}^{\mathrm{i} p_z q_0(t)/\hbar}
\:\mathrm{e}^{-\mathrm{i} m z \dot{q}_0(t)/\hbar}
\end{equation}
to transform the relevant lab-frame TDSE:
\begin{equation}
\mathrm{i}\hbar\:\frac{\partial}{\partial t}\Psi(\mathbf{r},t)=
\left[-\frac{\hbar^2\nabla^2}{2m}+U_\mathrm{F}(x,y,z-q_0(t))\right]\Psi(\mathbf{r},t) \:.
\end{equation}
As a result, the time-evolution for the relevant wave-function $\Phi(\mathbf{r},t)
\equiv\mathcal{U}\Psi(\mathbf{r},t)$ in the comoving frame is governed by another TDSE:
\begin{eqnarray}\label{eqSchroedingerequationTrapFrame}
\mathrm{i}\hbar\:\frac{\partial}{\partial t}\Phi(\mathbf{r},t) &=&
\Big[-\frac{\hbar^2\nabla^2}{2m}+\frac{m}{2}\:\dot{q}_0(t)^2 \nonumber \\
&+& m \ddot{q}_0\left(z + q_0\right)+U_\mathrm{F}(\mathbf{r})\Big]\Phi(\mathbf{r},t) \:.
\end{eqnarray}
The two terms $m\dot{q}_0(t)^2$ and $m \ddot{q}_0 q_0$ result only in time-dependent global phase factors
and can hereafter be safely neglected. One advantage of switching to the comoving frame is that we do not have
to compute the initial potential after every time step, but just the correction term $m\ddot{q}_0 q$ linear
in the acceleration of the potential minimum. In addition, we can restrict our ``simulation window'' around
the potential minimum, which obviates the need to take the whole expanded space of the transport process into
account.

Using the result from Eq.~\eqref{eqSchroedingerequationTrapFrame}, one time step in the FSOM for
our system can be written in the form
\begin{eqnarray}\label{eqFSOMFinal}
\Phi(\mathbf{r},t&+&\delta t)\approx\mathrm{exp}\left[-\frac{i}{\hbar}\:U_\mathrm{F}(\mathbf{r})
\frac{\delta t}{2}\right]\mathrm{F}^{-1}\Bigg[ \mathrm{exp}\left(-\mathrm{i}\frac{\hbar k^2}{2m}
\delta t\right) \nonumber \\
&\times& \mathrm{exp}\left[-\mathrm{i}\frac{k_z}{2}\delta \dot{q}_0(t)\delta t\right]
\mathrm{F}\Big(\mathrm{exp}\left[-\frac{i}{\hbar}z \delta \dot{q}_0(t)\delta t\right] \\
&\times& \mathrm{exp}\left[-\frac{i}{\hbar}\:U_\mathrm{F} (\mathbf{r}) \frac{\delta t}{2}\right]
\Phi(\mathbf{r},t)\Big)\Bigg] + O\left[(\delta t)^3\right] \nonumber \:,
\end{eqnarray}
with the velocity difference $\delta\dot{q}_0(t)\equiv\dot{q}_0(t+\delta t)-\dot{q}_0(t)$.
This equation is slightly more elaborate than Eq.~\eqref{eqFourierRepresentation}, because
the higher-order contributions resulting from the correction term $m\ddot{q}_0 q$ were already treated.
Therefore, the higher-order correction terms result solely from the application of the Baker-Campbell-Hausdorff
formula and the neglect of the time-ordered product. These terms depend on the commutators of the type
$\left[U_\mathrm{F}(\mathbf{r}),p^2\right]$, as well as the commutators $\left[H(t_1),H(t_2)\right]$
involving the Hamiltonian of the system at different times.

It should be stressed that by introducing a linear force of the form $F(t)=m\ddot{q}_0(t)$
in the lab frame, the resulting TDSE in the comoving frame [cf. Eq.~\eqref{eqSchroedingerequationTrapFrame}]
would not contain the term $ m \ddot{q}_0 q_0$. This is the so-called ``compensating-force approach'' and
results in the same TDSE as in the lab frame (up to global phase factors) and the ensuing perfect state transfer.
Yet, this method is much more challenging to implement experimentally for neutral atoms~\cite{Martinez-Garaot+:15}
and even impossible for systems containing trapped ions of more than one sort~\cite{Palmero+:14}.

\section{Results and Discussion} \label{ResultsDiscussion}
\subsection{Atom-transport fidelity: STA vs. eSTA}  \label{Fidelity_STA_eSTA}
\begin{figure}[b]
\includegraphics[width=0.875\linewidth]{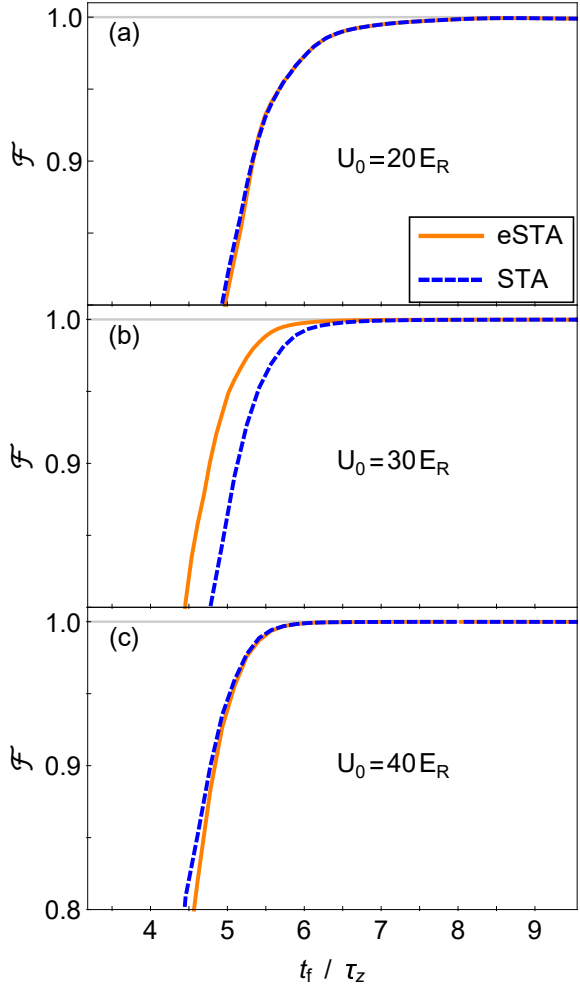}
\caption{\label{fig:fidelity2040}(Color online) The dependence of the atom-transport fidelity on the
transport time $t_f$, for a potential depth $U_0$ of (a) $20\:E_{\textrm{R}}$, (b) $30\:E_{\textrm{R}}$,
and (c) $40\:E_{\textrm{R}}$. The transport distance $d$ was set to $85\:l_z$,
while the transverse beam waists are $w_{x/y,0}=4.2\times 10^6\:l_z$.}
\end{figure}

\begin{figure}[t]
\includegraphics[width=0.875\linewidth]{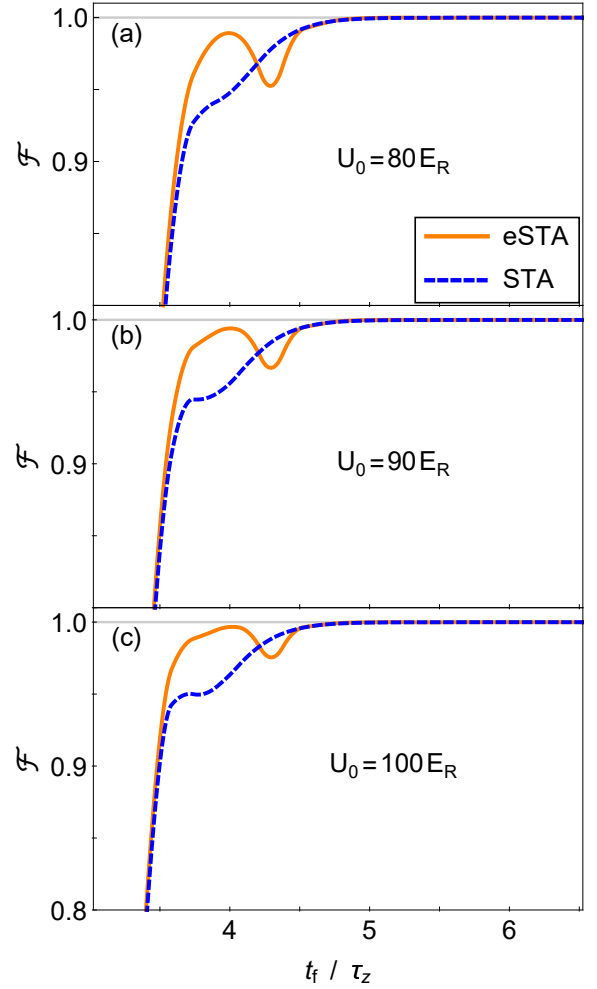}
\caption{\label{fig:fidelity5070}(Color online) The dependence of the atom-transport fidelity on
the transport time $t_f$, for a potential depth $U_0$ of (a) $80\:E_{\textrm{R}}$, (b) $90\:E_{\textrm{R}}$,
and (c) $100\:E_{\textrm{R}}$. The transport distance $d$ was set to $85\:l_z$,
while the transverse beam waists are $w_{x/y,0}=4.2\times 10^6\:l_z$.}
\end{figure}

\begin{figure}[t]
\includegraphics[width=0.875\linewidth]{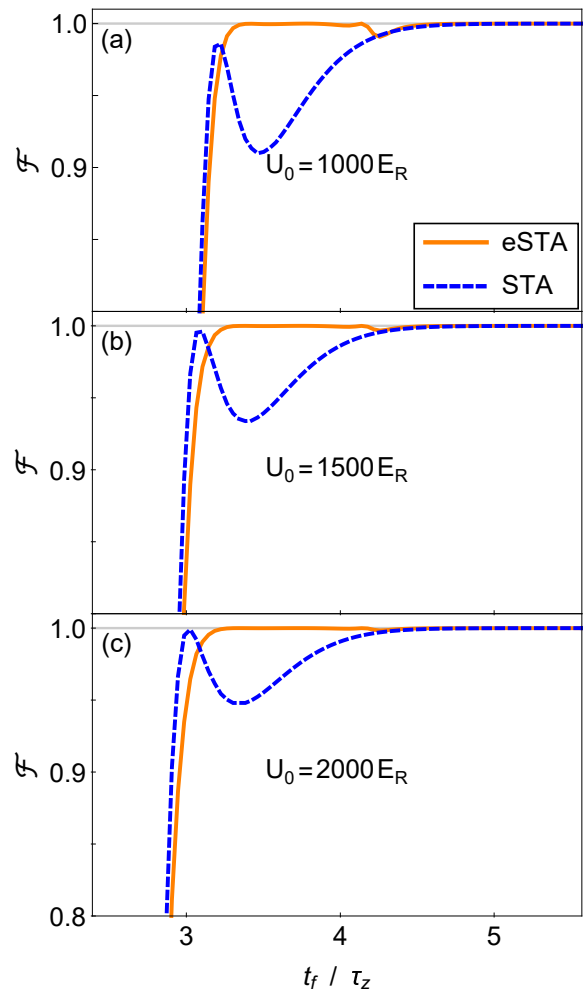}
\caption{\label{fig:fidelity80100}(Color online) The dependence of the atom-transport fidelity on
the transport time $t_f$, for a lattice depth $U_0$ of (a) $1000\:E_{\textrm{R}}$, (b) $1500\:E_{\textrm{R}}$,
and (c) $2000\:E_{\textrm{R}}$. The transport distance $d$ was set to $85\:l_z$, while the transverse
beam waists are $w_{x/y,0}=4.2\times 10^6\:l_z$.}
\end{figure}

In what follows, we present and analyze our results for the single-atom dynamics in an OCB, obtained
using the FSOM and the trap trajectories resulting from the STA and eSTA schemes [cf. Sec.~\ref{MovingTrap}].
The main figure of merit quantifying this process is the atom-transport fidelity $\mathcal{F}(t_f)=
|\braket{\Psi_{\textrm{target}}|\Psi(t_f)}|^2$, which is determined by the module of the overlap of
the target state $|\Psi_{\textrm{target}}\rangle$ (the ground state of the displaced OCB potential) and
the final atomic state $|\Psi(t_f)\rangle$. The dependence of the fidelity on the transport time $t_f$ is
illustrated for different optical-lattice depths $U_0$ in Figs.~\ref{fig:fidelity2040} -- \ref{fig:fidelity80100}.
These results correspond to the same transport distance $d=85\:l_z$, while the waists in
the transverse directions were set to $w_{x/y,0}=4.2\times 10^6\:l_z$.

One of the salient features of the obtained results is the collapse of the fidelity for short transport
times $t_f$, which is evident from Figs.~\ref{fig:fidelity2040} -- \ref{fig:fidelity80100}. This collapse
is, generally speaking, a consequence of the fact that the potential itself can only withstand an atomic
acceleration below a certain maximal value $|a_{\textrm{max}}|$ before the atom effectively escapes from
the trap and the corresponding fidelity drops rapidly. Namely, in the non-inertial reference frame that
moves with the atom the total lattice potential acquires an additional contribution that is linear in the
longitudinal coordinate, thus effectively leading to a tilted standing-wave potential in this accelerating
frame. As a result, the local minima of the standing wave dissapear completely for accelerations above
$|a_{\textrm{max}}|=U_0\:k/m$~\cite{Schrader+:01}. Being proportional to $U_0$, this maximal acceleration
becomes greater for deeper potentials~\cite{Hickman+Saffman:20}.

The collapse of the transport takes place when the maximal acceleration reached by an atom during the
transport process, which will be denoted by $|\tilde{a}_{\textrm{max}}|$ in the following, exceeds
$|a_{\textrm{max}}|$. While the lower bound on $|\tilde{a}_{\textrm{max}}|$ is quite generally given
by $2d/t_f^2$~\cite{Torrontegui+:11}, its actual value depends on the concrete chosen trap trajectory,
i.e. the path of the potential minimum. In particular, for the trajectory obtained using the STA
approach in Sec.~\ref{MovingTrap_STA} [cf. Eq.~\eqref{eqSolutionvectorPotential}], this value
is given by $|\tilde{a}_{\textrm{max}}|\approx 9.372\:d/t_f^2$. The fact that for a fixed transport
distance $|\tilde{a}_{\textrm{max}}|$ is inversely proportional to $t_f^2$ implies that for deeper
potentials (i.e. for a higher $|a_{\textrm{max}}|\propto U_0$) the actual maximal atomic acceleration
$|\tilde{a}_{\textrm{max}}|$ reaches the value of $|a_{\textrm{max}}|$ at shorter transport times $t_f$.
In other words, for deeper potentials the collapse of the fidelity takes place for shorter times $t_f$. This
is consistent with our numerical findings, illustrated in Figs.~\ref{fig:fidelity2040} -- \ref{fig:fidelity80100}
for gradually increasing potential depths. As can be inferred from these results, the characteristic transport
times $t_f$ pertaining to the occurrence of the collapse are around $4.5$, $3.4$, and $2.8$ $\tau_{\textrm{z}}$,
respectively, in Figs.~\ref{fig:fidelity2040} -- \ref{fig:fidelity80100} and clearly show the trend
of decreasing with the increase of the lattice depth.

Because for eSTA the modulations of the potential path through the optimization vector $\vec{\epsilon}$
are small contributions to the overall dynamics [cf. Sec.~\ref{MovingTrap_eSTA}], for deep-enough
lattices the collapse of the fidelity for eSTA-based atom transport takes place at approximately
the same transport times as for the corresponding STA scheme. However, it should be stressed that
for more shallow lattices (e.g. potential depths $U_0$ of $30$, $50$, and $60\:E_{\textrm{R}}$) the
transport time $t_f$ corresponding to the collapse can be notably different between STA and eSTA. This is due
to the fact that even small modulations (such as the modulation through the optimization vector
$\vec{\epsilon}$) can result in non-negligible differences between the maximal atomic accelerations
for STA and eSTA. This depends primarily on the modulation strength around the intermediate transport
times for which the maximal possible acceleration is exceeded in STA-based transport. On the other hand,
this also depends on the sign of the modulation, i.e. whether the modulation leads to higher or lower
atomic accelerations $|\tilde{a}_{\textrm{max}}|$.

Another interesting feature of the results obtained using the eSTA scheme is the slowly forming dip
for deeper potentials, as can be observed, e.g., in Fig.~\ref{fig:fidelity5070}(a) for
$t_f\approx 4.4\:\tau_{\textrm{z}}$.
The existence of this dip is a result of increasing transient excitation energies during the transport process
upon shortening the transport time $t_f$. Namely, as first discussed in Ref.~\cite{Torrontegui+:11},
the time-averaged transient excitation energy depends on $t_f$ according to $\bar{E}_\mrm{p,min}\propto t_f^{-4}$.
Consequently, the implications of the anharmonic character of potential become more and more prominent,
i.e. the assumption of the harmonic potential for the STA method begins to break down, resulting in
a worse performance of this method. At the same time, the performance of the eSTA approach becomes better
for deeper potentials and slowly approaches a perfect transport up until the aforementioned collapse of
the transport process. This performance improvement of the eSTA approach originates from the fact that
this approach relies on the smallness of the evolution parameters $\mu_S$ [cf. Sec.~\ref{MovingTrap_eSTA}],
which decreases with increasing potential depth.

\begin{figure}[t!]
\includegraphics[width=0.875\linewidth]{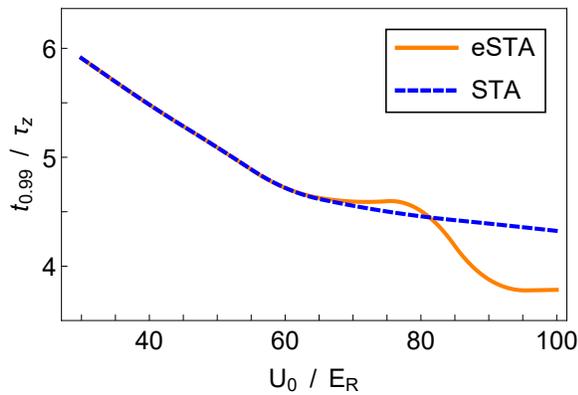}
\caption{\label{fig:pict099U}(Color online) The time $t_{0.99}$ for which a fidelity of $0.99$ is
first reached for different potential depths $U_0$. The transport distance $d$ was set to $85\:l_z$,
while the transverse beam waists are $w_{x/y,0}=4.2\times 10^6\:l_z$.}
\end{figure}

The relative efficiency of the eSTA-based atom transport  -- compared to its STA-based counterpart -- is
illustrated by Fig.~\ref{fig:pict099U}, which depicts the dependence of the time $t_{0.99}$ required to
first reach the fidelity of $0.99$ on the lattice depth $U_0$ for the fixed transport distance $d=85\:l_z$.
As already established, it can be inferred from this figure that the eSTA method constitutes an
improvement of the STA scheme for deeper potentials -- in this example for
$U_0\gtrsim 80\:E_{\textrm{R}}$ -- because a shorter transport time is needed to reach the same
fidelity of $0.99$. On the other hand, for more shallow potentials this is not the case. In fact,
in a narrow range between $U_0\sim 65\:E_{\textrm{R}}$ and $U_0\sim 80\:E_{\textrm{R}}$ eSTA even
yields results inferior to that of STA. This outcome -- that eSTA does not always result in higher
fidelities than STA -- seems to be consistent with the heuristic character of the eSTA approach.
However, eSTA is expected to reach perfect fidelity and be an improvement over STA for
$\mu_{\textrm{S}}\rightarrow 0$~[cf. Eq.~\eqref{eqCloseHamiltonian}].
\begin{figure}[b!]
\includegraphics[width=0.875\linewidth]{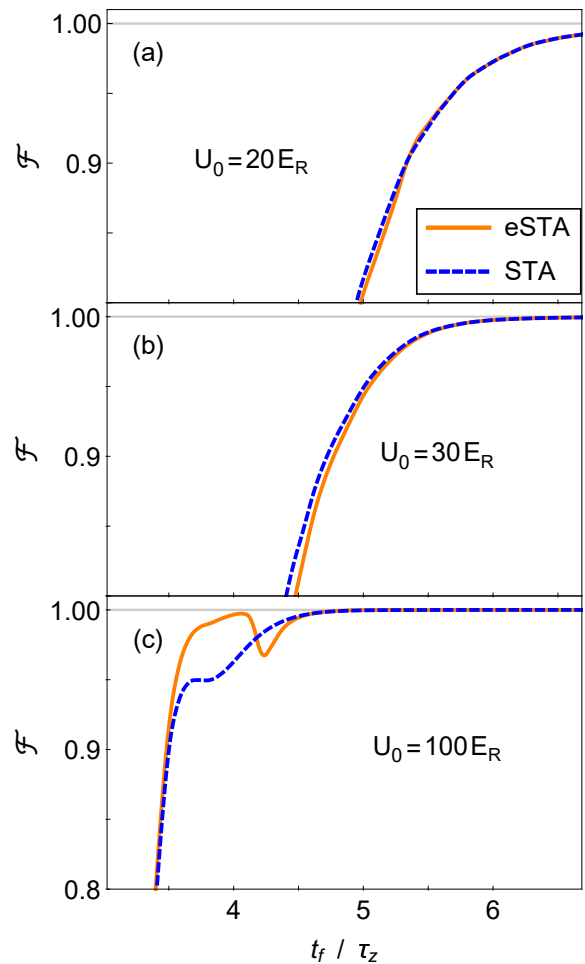}
\caption{\label{fig:pictWaistDep}(Color online) The dependence of the atom-transport fidelity on
the transport time $t_f$, for a lattice depth $U_0$ of (a) $20\:E_{\textrm{R}}$, (b) $30\:E_{\textrm{R}}$,
and (c) $100\:E_{\textrm{R}}$. The transverse beam waists are $w_{x/y,0}=300\:l_z$, while the transport
distance is $d=85\:l_z$.}
\end{figure}

It is pertinent to also comment on the obtained results for the atom-transport fidelity [cf. Figs.~\ref{fig:fidelity2040}
-- \ref{fig:fidelity80100}] from the point of view of the typical shapes of the corresponding trap-trajectory
solutions [cf. Figs.~\ref{fig:STAminimum} -- \ref{fig:eSTAminimum}]. What can be inferred is that the typical
times $t_f$ needed for a high-fidelity transport correspond to trap-trajectory solutions that do not display
oscillatory features. For instance, the eSTA trap trajectory for $t_f/\tau_z = 2$ in Fig.~\ref{fig:eSTAminimum}(a),
which has oscillating character, does not allow for a high-fidelity atom transport. In other words, in the
system at hand only non-oscillatory solutions for the trajectory of the moving lattice can enable such transport.

For the sake of completeness, it is worthwhile to briefly discuss the effect of varying transverse beam
waists on the efficiency of atomic transport. Our calculations show that the variation of the waists leads
to appreciable changes (for fixed values of other relevant parameters) of the fidelity only for rather
shallow lattices, i.e. for lattice depths as small as several tens of $E_{\textrm{R}}$. For larger lattice
depths the results are practically insensitive to the size of the transverse beam waists. This is illustrated
in Fig.~\ref{fig:pictWaistDep}, where the dependence of the fidelity on the transport time is shown for
the lattice depths $U_0$ of $20,\:30$, and $100\:E_{\textrm{R}}$ with the relevant waists chosen to be
$w_{x/y,0}=300\:l_z$. For the parameter choice corresponding to Fig.~\ref{fig:pictWaistDep} the behavior
of the fidelity changes for lattice depths $U_0$ just slightly above $30\:E_{\textrm{R}}$ and remains
essentially unchanged upon further increase of $w_{x/y,0}$.

\subsection{Comparison to other approaches} \label{SineAndTriangleComp}
In what follows, we complement our analysis of STA and eSTA results for the atom-transport fidelity
by comparing these results to those originating from other known approaches. To be more precise, we
consider approaches based on the use of sine-shaped and triangular velocity profiles for the potential
path. The time-dependent forms of these two profiles are given by:
\begin{align}
q_0^\mathrm{ s } (t)
&
= \frac{ v_0 }{ 2 }\left[t - \frac{\sin\left( 2 \pi \frac{t}
{ t_f}\right)}{2 \pi } t_f \right],
\\
q_0^\mathrm{ t } (t)
&
=
\begin{cases}
v_0 \, t^2 / t_f, & \text{for} \; \; 0 \leq t \leq t_f/2,\\
v_0  \left(2 \, t -  t_f /2 - t^2 / t_f \right), & \text{for}
\; \; t_f /2 < t \leq t_f  \:,
\end{cases}
\end{align}
with the maximal velocity $v_0 = 2d/t_f$.

The approach based on the triangular velocity profile is also known as the {\em bang-bang}
approach~\cite{Torrontegui+:11,Chen+:11,Ding+:20}. As a consequence of discontinuities in its corresponding
acceleration profile, this approach
leads to additional motional heating in the regime of fast transport. As a result, it showed a relatively
poor performance in some previous studies, e.g. in Ref.~\cite{Hickman+Saffman:20}. On the other hand, the
sine-shaped profile represent an improvement over the bang-bang approach, since its attendant acceleration
is continuous during the entire transport process. However, it is plausible to expect that STA and eSTA
approaches should lead to much better results than these pre-selected velocity profiles. Namely, the STA
approach is based upon inverse engineering and makes use of the specific form of the Hamiltonian in question
to obtain a tailored trap trajectory. Likewise, being based on STA solutions for simplified systems, eSTA
solutions inherit this last property of their STA counterparts.

\begin{figure}[t]
\includegraphics[width=0.875\linewidth]{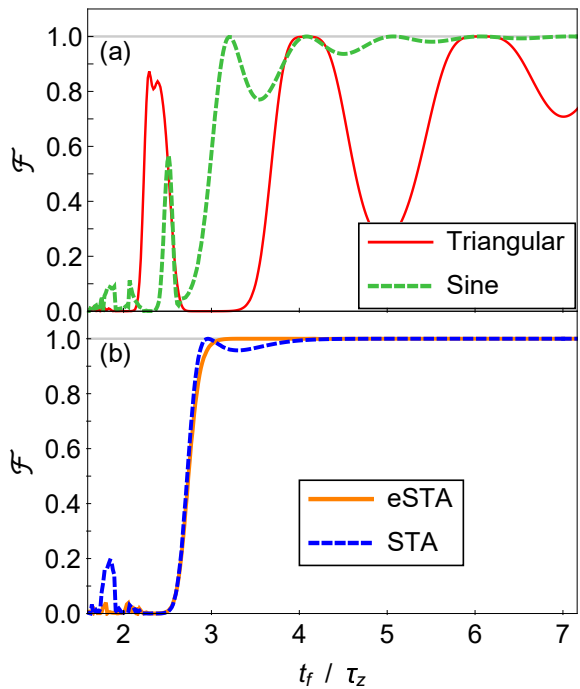}
\caption{\label{fig:fidelityComparison}(Color online) The atom transport fidelity for an atom prepared in
the longitudinal motional ground state and moved by a distance $d=140\:l_z$. The results correspond to (a)
sine-shaped and triangular velocity profiles and (b) STA and eSTA methods. The lattice depth is set to $U_{0}
= 2610\:E_{\textrm{R}}$, while the transverse waists are $w_{x,0}=1790\:l_z$ and $w_{y,0}=537\:l_z$.}
\end{figure}

The fidelities obtained using triangular and sine-shaped velocity profiles are compared to those resulting
from the application of STA/eSTA methods in Fig.~\ref{fig:fidelityComparison}. The plot shows the dependence
of $\mathcal{F}$ on the transport times $t_f$ for an atom that is initially prepared in the longitudinal
motional ground state and moved by a fixed distance (here $d=140\:l_z$), with the target state being the
ground state of the displaced OCB potential.

The triangular velocity profile shows strong oscillations in fidelity, a trend that gradually becomes
more prominent upon reducing transport times $t_f$, up until the complete breakdown of the
fidelity for $t_f\approx 2.1\:\tau_{\textrm{z}}$ [cf. Fig.~\ref{fig:fidelityComparison}(a)]. Somewhat
better results are obtained for the sine-shaped velocity profile. Even though the latter also show
oscillations, these are much less pronounced than in the triangular case and start for much shorter
transport times; the complete breakdown occurs for $t_f\approx 2.3\:\tau_{\textrm{z}}$.

In accordance with the aforementioned expectation, a significant improvement over these previous results is
obtained using STA and eSTA, where only one major drop in fidelity takes place for $t_f\approx 2.9\:\tau_{\textrm{z}}$
[cf. Fig.~\ref{fig:fidelityComparison}(b)]. While eSTA results in smaller fidelities than STA for times
$t_f\approx 2.9\:\tau_{\textrm{z}}$ very close to the breakdown point, the eSTA method still leads to
slightly larger fidelities than STA for almost all transport times.

\section{Summary and Conclusions} \label{SummaryConclusions}
In summary, using a combination of advanced analytical and numerical techniques in this paper
we investigated fast single-atom transport in moving optical lattices (optical conveyor belts).
Unlike previous theoretical studies of fast atomic transport, which were almost exclusively
based on simplified scenarios -- such as strictly one-dimensional systems and/or purely
harmonic trapping potentials -- we studied this phenomenon by taking fully into account
the three-dimensional, anharmonic trapping potential of the system under consideration.

Our results for atom-transport fidelities -- obtained using both STA and eSTA
approaches -- correspond to realistic values of the relevant system parameters
(beam waists, lattice depths, transport distances, etc.).
Moreover, our study demonstrates the feasibility of applying the recently proposed eSTA
method to a realistic experimental system. It shows that eSTA -- envisioned as an
improvement of the existing STA techniques -- indeed yields more efficient atom transport
in optical conveyor belts than STA in a broad range of system parameters.

It can be expected that our present study will motivate further attempts towards realistic
modelling of single-atom transport in various optically-trapped atomic systems, such as
optical lattices of different geometry~\cite{Stojanovic+:08,Hofer+:12}. In addition, while
in the present work only near-ground state atoms have been considered, it is worthwhile to
also investigate the finite-temperature effects (leading, e.g., to finite atom lifetime
in traps) and optically-induced heating (due to optical-potential fluctuations). Likewise,
this study is of utmost relevance for future experiments in optical conveyor belts. In
particular, an experimental corroboration of our results for the atom-transport fidelities
is clearly called for.

\begin{acknowledgments}
This research was supported by the Deutsche Forschungsgemeinschaft
(DFG) -- SFB 1119 -- 236615297.
\end{acknowledgments}

\onecolumngrid

\appendix
\section{Derivation of the expression for $G_n$} \label{derivationG_n}
In the following we derive an expression that can be used for the numerical evaluation of the first
auxiliary function $G_{\mathbf{n}}$ [cf. Eq.~\eqref{eqExpressionG}] in our problem. For the sake of
brevity, the multi-indices $\mathbf{n}\equiv (n_x,n_y,n_z)$ and $\mathbf{n}_r\equiv (n_x,n_y)$ are used.
It should be borne in mind that the main quantum number $n$ of a 3D harmonic oscillator is given by
the sum of the quantum numbers of three 1D oscillators, i.e. $n=n_x+n_y+n_z$.

By inserting the transport modes of a 3D harmonic Hamiltonian, written in the coordinate representation,
into Eq.~\eqref{eqExpressionG} we obtain the expression
\begin{equation}\label{eqGStart}
\begin{split}
G_{\mathbf{n}} =
& - \, \int_0^{t_f} dt \int_{-\infty}^{\infty} dZ \int_{-\infty}^{\infty} dY \int_{-\infty}^{\infty}
dX \,  \frac{\mrm{ exp } \left[ \mrm{i} \left( \omega_x n_x + \omega_y n_y + \omega_\mrm{z} n_z\right)
t \right]}{\sqrt{2^n n_x! n_y! n_z! \pi^3}} \,
\\
&
\times  \mrm{H}_{n_x} \left( X \right) \mrm{H}_{n_y} \left( Y \right) \mrm{H}_{n_z} \left[ Z_C(t) \right]
\, \mrm{exp} \left( - X^2 \right) \mrm{exp} \left( - Y^2 \right) \mrm{exp} \left[ - Z_C(t)^2 \right]
\\
&
\times \Bigg( \frac{C\,P_0 \, \mrm{cos} \left[\sqrt{2} \, k \, l_z \, \mn{Z}_0(t)\right]^2}{w_x\left[ \mn{Z}_0(t)
\right] w_y\left[ \mn{Z}_0(t) \right]} \,
\mrm{exp} \left[ -4 \left( \frac{\mn{X}^2 \, l_\mrm{x}^2}{w_x\left[ \mn{Z}_0(t) \right]^2} \, + \;
\frac{\mn{Y}^2 \, l_\mrm{y}^2}{w_y\left[ \mn{Z}_0(t) \right]^2} \right) \right]
\\
&
\mrm{ +\frac{\hbar}{2}\left[ \omega_x \mn{X}^2 + \omega_y \mn{Y}^2 + \omega_\mrm{z} \mn{Z}_0(t)^2
\right]}- U_0\Bigg)\:,
\end{split}
\end{equation}
with the dimensionless coordinates $X = x / ( l_\mrm{x}\sqrt{2})$, $Y= y / ( l_\mrm{y}\sqrt{2})$, $Z = z / ( l_z \sqrt{2})$,
and new functions $ Z_0(t)\equiv Z -  q_0(t) / ( l_z\sqrt{2}) $, $ Z_C(t)\equiv Z -  q_{c,z}(t) / (l_z\sqrt{2}) $.
For notational convenience, the waists will hereafter be denoted by $w_{x/y}\left[ \mn{Z}_0(t) \right]$,
instead of $w_{x/y}\left[ \mn{Z}_0(t) \,l_z \, \sqrt{2}\right]$. The final expression for $G_{\mathbf{n}}$
will be obtained by treating the different terms and integrations separately from each other.

\subsection{Integration over the transverse directions}\label{SubSubSecIntegrationXY}
We first carry out the integrations in $X$ and $Y$ directions, because those are conceptually easier
to do than the $Z$ integration. Therefore, the integral we are considering here is given by
\begin{equation}\label{eqGRIntegration}
\begin{split}
\mrm{I}_r^{\mathbf{n}_r} \left[ Z_0(t) \right] & = \int_{-\infty}^{\infty} dY \int_{-\infty}^{\infty} dX
\, \mrm{H}_{n_x} \left( X \right) \mrm{H}_{n_y} \left( Y \right) \mrm{exp} \left( - X^2 \right) \mrm{exp}
\left( - Y^2 \right)
\\
&
\hspace{0.5cm} \times
\Bigg( A\left[Z_0(t)\right] \, \mrm{exp}\left[ -4 \left( \frac{X^2\, l_x^2}{w_x\left[ Z_0(t) \right]^2} \,
+ \, \frac{Y^2\, l_y^2}{w_y\left[ Z_0(t) \right]^2} \right) \right]
\\
&
\hspace{0.5cm}
+\frac{\hbar}{2}\left[ \omega_x X^2 + \omega_y Y^2 + \omega_z Z_0(t)^2 \right] - U_0 \Bigg) \:,
\end{split}
\end{equation}
with $A(Z) = CP_0\mrm{cos}^2\left(k \, l_z \, Z \right)/[w_x \left( Z \right) w_y \left( Z \right)]$
and $U_0 = CP_0/(w_{x,0} w_{y,0})$.
Using the orthogonality and the recurrence relation of Hermite polynomials~\cite{ChowBOOK:00}
\begin{align}
\label{eqHermiteOrthogonal}
\int_{-\infty}^\infty d x \, & \mrm{H}_m( x ) \, \mrm{H}_l( x ) \, \mrm{exp}
\left( -x^2 \right) = \sqrt{\pi} \, 2^m \, m! \, \delta_{l,m}  \:,
\\
\label{eqRelationHermite1}
x & \mrm{H}_m(x)=\frac{1}{2}\mrm{H}_{m+1}(x) + m \mrm{H}_{m-1}(x) \:,
\end{align}
the $X$-integration of the second term in Eq.~\eqref{eqGRIntegration} can readily be carried out.
It yields the following result:
\begin{equation}\label{eqGRIntegration2x}
\begin{split}
\mrm{I}_{r,2}^{n_x}\left[ Z_0(t) \right]
&
= \frac{\hbar}{2} \int_{-\infty}^{\infty} d X \, \left[ \mrm{ \omega_x \mn{X}^2 +
\omega_y \mn{Y}^2 + \omega_\mrm{z}^2 \mn{Z}_0(t)^2 } - \frac{2 U_0 }{\hbar}\right] \,
\mrm{H}_{n_x} \left( X \right)  \mrm{exp} \left( - X^2 \right)
\\
&
=
\frac{\hbar \omega_x}{4} \sqrt{\pi} \left(  \delta_{n_x,0} + 4 \, \delta_{n_x,2} \right)
+ \sqrt{\pi}\frac{\hbar}{2} \left[  \omega_y Y^2 +  \omega_z Z_0(t)^2 -\frac{2U_0}{\hbar}
\right] \delta_{n_x,0} \:.
\end{split}
\end{equation}

Owing to the symmetry of the problem, the $Y$-integration of Eq.~\eqref{eqGRIntegration2x} is conceptually
equivalent to the $X$-integration. Therefore, we just state the final result for the integrated second term
[cf. Eq.~\eqref{eqGRIntegration}]:
\begin{equation}\label{eqGRIntegration2Final}
\begin{split}
\mrm{I}_{r,2}^{\mathbf{n}_r}\left[ Z_0(t) \right]
& = \pi \frac{\hbar}{2} \omega_z \, Z_0(t)^2 \delta_{n_x,0} \, \delta_{n_y,0}
\\
&
\hspace{0.5cm}
+ \pi \hbar \left[\left(\frac{\omega_x + \omega_y}{4}  -\frac{U_0}{\hbar}\right) \delta_{n_x,0} \,
\delta_{n_y,0}  +  \omega_x \, \delta_{n_x,2} \, \delta_{n_y,0} + \omega_y \, \delta_{n_x,0} \,
\delta_{n_y,2} \right]\:.
\end{split}
\end{equation}

Let us now focus on the first term in Eq.~\eqref{eqGRIntegration}. We can restrict our calculations
to the integration of
\begin{equation}\label{eqGRIntegration1}
\mrm{I}_{x,1}^{n_x}\left[ Z_0(t) \right] = \int_{-\infty}^{\infty} d X \, \mrm{H}_{n_x}
\left( X \right) \mrm{exp} \left( - X^2 \right) \, \mrm{exp} \left( -4\frac{ X^2\, l_\mrm{x}^2}
{w_x \left[ Z_0(t) \right]^2} \right) \:.
\end{equation}

As a consequence of the presence of the second exponential in the last integral, we cannot simply use the
orthogonality relation \eqref{eqHermiteOrthogonal} of Hermite polynomials to evaluate it. However, this
integral can be computed using the formula of Fa\'{a} di Bruno~\cite{Weisstein:21} for Hermite polynomials
\begin{equation}\label{eqHermiteFaadibruno}
\mrm{H}_m(x)=(-1)^m\sum_{k_1+2k_2=m} \frac{m!}{k_1! k_2!} (-1)^{k_1 + k_2}
\left( 2 \mn{x} \right)^{k_1}.
\end{equation}
In addition, we make use of the identity
\begin{equation}\label{eqIntegralRelation}
\int_{-\infty}^\infty d x \, x^n \, \mrm{exp} \left( - \mrm{a} x^2 + \mrm{b} x +\mrm{c} \right)
= \mrm{exp \left( \frac{b^2}{4a} + c \right)} \sum_{k=0}^{\lfloor n/2 \rfloor}
\begin{pmatrix}
n \\
2k
\end{pmatrix}
\left(\mrm{\frac{b}{2a}}\right)^{n-2k}
\frac{ \Gamma \left( k + 1/2 \right) }{\mrm{a}^{k + 1/2}} \:,
\end{equation}
where $\Gamma(x)$ is the gamma function. Putting everything together, the following result is finally obtained:
\begin{equation}\label{eqGeneralGRIntegration}
\begin{split}
\mrm{I}_{x,1}^{n_x}\left[ Z_0(t) \right]
&
= \sum_{k_1+2k_2=n_x} \frac{n_x!}{k_1! k_2!} (-1)^{n_x+k_1 + k_2}
\int_{-\infty}^\infty \mn{d}X\left( 2 X \right)^{k_1} \mrm{exp} \left( -X^2 \right) \,
\mrm{exp} \left( -4\frac{X^2\, l_\mrm{x}^2}{w_x\left[Z_0(t)\right]^2} \right)
\\
&
=\sum_{ \substack{k_1 + 2 k_2 = n_x \\ k_1 \mrm{even} } } \frac{n_x!}{k_1! k_2!} (-1)^{k_2}
\Gamma \left( \frac{k_1+1}{2} \right) \:.
\end{split}
\end{equation}
In the last step we made use of the fact that the integral in Eq.~\eqref{eqGRIntegration1} is equal to
zero for odd values of $n_x$ due to the symmetry of the integrand. Similar result can also be obtained
for the $Y$-integration. Thus, the final integrated form for the first term in Eq.~\eqref{eqGRIntegration},
up to the $Z$-dependent factor $A\left[Z_0(t)\right]$, is given by
\begin{equation}\label{eqGRIntegration1final}
\begin{split}
\mrm{I}_{r,1}^{\mathbf{n}_r}\left[ Z_0(t) \right] =\sum_{ \substack{k_1 + 2 k_2 = n_x \\ k_1 \mrm{even} } }
\sum_{ \substack{\tilde{k}_1 + 2 \tilde{k}_2 = n_y \\ \tilde{k}_1 \mrm{even} } }
&
\frac{n_x! \, n_y!}{k_1! k_2! \, \tilde{k}_1! \tilde{k}_2!} (-1)^{k_2 + \tilde{k}_2}
\, 2^{k_1+\tilde{k}_1} \,
\left( \frac{w_x\left[Z_0(t)\right]^2}{4 \, l_\mrm{x}^2 + w_x\left[Z_0(t)\right]^2}\right)^{\frac{k_1 + 1}{2}}
\\
&
\hspace{0.35cm} \times
\Gamma \left( \frac{k_1+1}{2} \right) \,
\Gamma \left( \frac{\tilde{k}_1+1}{2} \right)
\left( \frac{w_y\left[Z_0(t)\right]^2}{4 \, l_\mrm{y}^2 +  w_y\left[Z_0(t)\right]^2} \right)
^{\frac{\tilde{k}_1 + 1}{2}}.
\end{split}
\end{equation}

\subsection{Integration over the longitudinal direction}\label{SubSubSecIntegrationZ}
The most general form we can obtain for $G_{\mathbf{n}}$ after the integrations over the
$X$- and $Y$ coordinates is given by
\begin{equation}\label{eqGZ}
\begin{split}
G_{\mathbf{n}}  = & - \, \int_0^{t_f} dt \int_{-\infty}^{\infty} dZ \frac{ \mrm{ exp } \left[ \mrm{i}
\left( \omega_x n_x + \omega_y n_y + \omega_\mrm{z} n_z\right) t \right]}{\sqrt{2^n n_x! n_y! n_z!\pi}}
\\
&
\times \mrm{H}_{n_z} \left[ Z_C(t) \right] \, \mrm{exp} \left[ - Z_C(t)^2 \right]
\Bigg( \frac{1}{\pi} \, A\left[Z_0(t)\right] \, I_{r,1}^{\mathbf{n}_r}[Z_0(t)]
+ \frac{\hbar}{2} \omega_\mrm{z} Z_0(t)^2 \delta_{n_x,0} \, \delta_{n_y,0}
\\
&
+ \hbar \left[\left(\frac{\omega_x+\omega_y}{4} -\frac{U_0}{\hbar} \right) \, \delta_{n_x,0} \,
\delta_{n_y,0} +  \omega_x \delta_{n_x,2} \, \delta_{n_y,0}+ \omega_y \, \delta_{n_x,0} \, \delta_{n_y,2}
\right] \Bigg).
\end{split}
\end{equation}
For the $Z$-integration we treat the three terms in the brackets of Eq.~\eqref{eqGZ} independently.

The integration of the third term is conceptually the simplest one and is thus treated first.
The integral we need to evaluate has the form
\begin{equation}\label{eqGIntegrationZ3}
\mrm{I}_{z,3}^{\mathbf{n}} = B^{\mathbf{n}_r} \int_{-\infty}^{\infty} dz \, ( l_z \, \sqrt{2} )^{-1}
\, \mrm{H}_{n_z} \left[ Z_C(t) \right] \mrm{exp} \left[ - Z_C(t)^2 \right],
\end{equation}
where we have set $B^{\mathbf{n}_r} \equiv \hbar \left[\left(\omega_x/4 + \omega_y/4 -
U_0/\hbar  \right) \delta_{n_x,0} \, \delta_{n_y,0} + \omega_x \, \delta_{n_x,2} \,
\delta_{n_y,0} + \omega_y \, \delta_{n_x,0} \, \delta_{n_y,2} \right]$.

Since we want to use orthogonality relation \eqref{eqHermiteOrthogonal},
we have to rewrite the Hermite polynomial and the exponential function
such that their arguments become independent of $q_0(t)$ and $q_{c,z}(t)$.
We can accomplish this using the relations of the generating function of
Hermite polynomials and the following sum representation~\cite{ChowBOOK:00}
\begin{align}
\label{eqHermiteRelations1}
\mrm{ exp \left( 2 \mn{x t} - \mn{t}^2 \right)} &= \mrm{\sum_{\mn{m}=0}^\infty
\frac{H_\mn{m}(\mn{x}) \mn{t^m}}{\mn{m}!} } \:,
\\
\label{eqHermiteRelations2}
\mrm{H}_m(x+y) & =  \sum_{k=0}^m
\begin{pmatrix}
\mrm{m} \\
k
\end{pmatrix}
\mrm{H}_k(x)\left( 2 y \right)^{m-k}.
\end{align}

We are now able to calculate the integral and obtain the following
\begin{equation}\label{eqGIntegrationZ32}
\begin{split}
\mrm{I}_{z,3}^{\mathbf{n}} & =  B^{\mathbf{n}_r} \int_{-\infty}^{\infty} d Z \, \sum_{k=0}^{n_z}
\begin{pmatrix}
n_z\\
k
\end{pmatrix}
\mrm{H}_{k} \left( Z  \right) \left[ -  \sqrt{2} \frac{  q_{c,z}(t) }{ l_z } \right]^{n_z-k}
\\
&
\hspace{0.5cm} \times \mrm{exp} \left( - Z^2 \right) \sum_{m=0}^\infty \frac{ \mrm{H}_m \left(
Z \right)  }{ m! } \, \left[ \frac{ q_{c,z}(t) }{ \sqrt{2} \,l_z } \right]^m
\\%
&
= B^{\mathbf{n}_r} \sqrt{\pi} \left[ \sqrt{2} \frac{  q_{c,z}(t) }{ l_z } \right]^{n_z} \sum_{k=0}^{n_z}
\begin{pmatrix}
n_z \\
k
\end{pmatrix}
(-1)^{n_z-k}.
\end{split}
\end{equation}
The last line of Eq.~\eqref{eqGIntegrationZ32} vanishes for any value of $n_z$ except for $n_z=0$,
which can be seen by making use of the binomial theorem. Hence, the solution to the first integral
in $Z$ is given by the simple form
\begin{equation}\label{eqGIntegrationZFinal}
\mrm{I}_{z,3}^{\mathbf{n}} = B^{\mathbf{n}_r} \sqrt{\pi} \, \delta_{n_z,0}\:.
\end{equation}

The $Z$-integration of the second term of Eq.~\eqref{eqGZ} can be carried out by
combining the above integration steps and using relation \eqref{eqRelationHermite1}.
Thus, the second integral in $Z$ is given by
\begin{equation}\label{eqGInetgrationZ2}
\begin{split}
\mrm{I}_{z,2}^{\mathbf{n}}(t)
&
= \frac{\hbar}{2} \omega_\mrm{z} \int_{-\infty}^{\infty} dZ \, \mrm{exp} \left[ - Z_C(t)^2 \right]
\, Z_0(t)^2 \, \mrm{H}_{n_z} \left[ Z_C(t) \right] \delta_{n_x,0} \, \delta_{n_y,0}
\\
&
= \frac{\hbar}{2} \omega_\mrm{z} \int_{-\infty}^\infty dZ \sum_{m=0}^\infty \frac{1}{m!} \, \left[ \frac{ q_{c,z}(t) }{ \sqrt{2} \,l_z } \right]^m \, \mrm{exp} \left[ -Z^2 \right] \, \delta_{n_x,0} \, \delta_{n_y,0}
\Bigg[\frac{1}{4} \, \mrm{H}_{m + 2} \left( Z \right)
\\
&
\hspace{0.5cm} + \left(m + \frac{1}{2} \right) \mrm{H}_{m} \left( Z \right) + m \, (m - 1) \mrm{H}_{m - 2}
\left( Z \right)
+ \frac{q_0(t)^2}{2 \,l_z^2} \, \mrm{H}_{m} \left( Z \right)
\\
&
\hspace{0.5cm} - \sqrt{2}  \, \frac{q_0(t)}{l_z} \left[\frac{1}{2} \mrm{H}_{m+1} \left( Z \right) + m \,
\mrm{H}_{m - 1} \left( Z \right) \right] \Bigg]
\sum_{l=0}^{n_z}
\begin{pmatrix}
n_z\\
l
\end{pmatrix}
\mrm{H}_{l} \left( Z \right) \left[- \sqrt{2} \frac{q_{c,z}(t)}{l_z} \right]^{n_z-l}
\\
&
= \frac{\hbar}{2} \, \sqrt{\pi} \, \omega_\mrm{z} \, \delta_{n_x,0} \, \delta_{n_y,0} \left[ \sqrt{2} \,
\frac{q_{c,z}(t)}{l_z} \right]^{n_z} \,
\sum_{l=0}^{n_z}
\begin{pmatrix}
n_z\\
l
\end{pmatrix}
(-1)^{n_z - l}
\Bigg[
\frac{l\, (l-1)}{2} \, \left[ \frac{q_{c,z}(t)}{l_z} \right]^{-2}
- l  \frac{ q_0(t) }{ q_{c,z}(t) }
+ l
\Bigg]\:,
\end{split}
\end{equation}
In the last step we utilized the orthogonality of Hermite polynomials [cf. Eq.~\eqref{eqHermiteOrthogonal}],
the binomial theorem and the general condition $n>0$. Using mathematical induction, we can further simplify
this last result and obtain the following form:
\begin{equation}
\mrm{I}_{z,2}^{\mathbf{n}}(t)= \frac{\hbar \, \omega_\mrm{z}}{ \sqrt{2} \, l_z } \, \sqrt{\pi}  \, \delta_{n_x,0} \, \delta_{n_y,0}
\Big(\delta_{n_z,1} \left[q_{c,z}(t)-q_0(t)\right] + \delta_{n_z,2}\Big)\:.
\end{equation}
This shows that the only nonvanishing contributions are those with $n_z=1,2$.

The last integral that we have to compute corresponds to the first term in Eq.~\eqref{eqGZ} and has the form
\begin{equation}\label{eqIntegrationZ1}
\mrm{I}_{z,1}^{\mathbf{n}}(t)
=  \frac{C \, P_0}{\pi}\, \int_{-\infty}^{\infty} dZ \,  \frac{ I_{r,1}^{\mathbf{n}_r}[Z_0(t)] \, \mrm{H}_{n_z}
\left[ Z_C(t) \right] \, \mrm{exp} \left[ - Z_C(t)^2 \right]}{w_x\left[ \mn{Z}_0(t)\right]w_y\left[ \mn{Z}_0(t)
\right]} \cos^2\left[\sqrt{2} \, k \, l_z Z_0(t)\right] \:.
\end{equation}
The dependence of the denominator on $Z$ makes it impossible to find an analytical solution for the
above integral even for concrete values of $n_z$. Therefore, as part of our optimization procedure,
we perform numerical evaluation of this integral.

Putting the results of the last two subsections together, we obtain the integrated form of $G_{\mathbf{n}}$:
\begin{equation}\label{eqGIntegrationT}
\begin{split}
G_{\mathbf{n}} & = -\int_0^{t_f} dt \:\frac{ \mrm{exp}\left[ \mrm{i} \left( \omega_x n_x +\omega_y n_y +
\omega_\mrm{z} n_z \right) t \right] }{\sqrt{2^n n_x! n_y! n_z!}} \\
& \hspace{0.5cm} \times \left[  \hbar \delta_{n_z,0}\left( \omega_x \, \delta_{n_x,2}\delta_{n_y,0}  +
\omega_y \, \delta_{n_x,0}\delta_{n_y,2} \right) + \frac{1}{\sqrt{\pi}} \mrm{I}_{z,2}^{\mathbf{n}}(t) +
\frac{1}{\sqrt{\pi}} \mrm{I}_{z,1}^{\mathbf{n}}(t) \right].
\end{split}
\end{equation}
Because an analytic solution for $\mrm{I}_{z,1}^{\mathbf{n}}(t)$ does not exist, this integral can only
be computed numerically.
\subsection{Approximation for $G_{\mathbf{n}}$}\label{SubSubSecApproxGn}
Because the numerical evaluation of two-dimensional integrals can be rather time-consuming,
we simplify the $z$-dependent denominator within the function $\mrm{I}_{z,1}^{\mathbf{n}}(t)$.
By analyzing the exponential function in Eq.~\eqref{eqIntegrationZ1}, we see that the main
contributions of $\mrm{I}_{z,1}^{\mathbf{n}}(t)$ are localized around the classical path of
the particle $q_{c,z}(t)$. Thus, we will first express the argument of the $z$-dependent
denominator in terms of $q_{c,z}(t)$, resulting in
\begin{equation}\label{eqDenominatorApprox1}
\begin{split}
\sqrt{2} \, Z_0(t) \, l_z
&
= q_{c,z}(t) +|\ddot{q}_{c,z}(t)/\omega_\mrm{z}^2|
\leq q_{c,z}(t) +\frac{\ddot{q}_{c,z}^\mrm{max}}{\omega_\mrm{z}^2}
\\
&
\approx q_{c,z}(t) +\frac{10 d}{t_f^2\omega_\mrm{z}^2}
\leq q_{c,z}(t) +\frac{1}{\sqrt{2}} \tilde{U}^{7/4} l_z \:.
\end{split}
\end{equation}
It should be borne in mind that within the STA solution, the acceleration of the particle $\ddot{q}_{c,z}(t)$
is connected to the difference in classical particle and potential paths [cf. Eq.~\eqref{eqForcedHarmonicOscillator}].
Moreover, in the second line the following inequality for the particle acceleration was used:
\begin{equation}\label{eqParticleAcceleration}
\begin{split}
\ddot{q}_{c,z}^\mrm{max} := \underset{t\in \left[ 0 , t_\mrm{f} \right]}{\mrm{max}}\ddot{q}_{c,z}(t)
= \frac{d}{t_f^2}\tilde{a} \leq |a_\mrm{max}| = \frac{1}{\sqrt{2}} \tilde{U}^{7/4}\frac{l_z}{\omega_z^{-2}} \:,
\end{split}
\end{equation}
where the dimensionless lattice depth $\tilde{U}\equiv U_0/E_{\textrm{R}}$ was introduced.
In Eq.~\eqref{eqParticleAcceleration} $|a_\mrm{max}|$ is the maximal acceleration of the trap (i.e., that
of the moving OCB potential), while the dimensionless parameter $\tilde{a}$ is the maximal acceleration
of an atom in the trap expressed in units of $d/t_f^2$. As already stated in Sec.~\ref{Fidelity_STA_eSTA},
the upper bound for $\tilde{a}$ is close to $10$, more precisely $9.372$, while the lower bound equals
$2$~\cite{Torrontegui+:11}.

Now, let us examine the Rayleigh lengths by looking at the following expression
\begin{equation}\label{eqApproxDenominator}
\begin{split}
\frac{ Z_{R,x/y} }{ \sqrt{2} \, l_z } & = \left[ \frac{m^2}{\hbar^2}\frac{U_0}{m} \left( \frac{1}{Z_{R,x}^2} + \frac{1}{Z_{R,y}^2}
+ 2 k^2\right)\right]^{1/4} Z_{R,x/y}\\
&
\gg \left(\frac{U_0 m}{\hbar^2 } 4 k^2 \right)^{1/4} k^{-1} = \left(2\tilde{U}\right)^{1/4} \:,
\end{split}
\end{equation}
with the characteristic length-scale $l_z=\sqrt{\hbar/(2 m \omega_z)}$, the frequency
$\omega_\mrm{z} = \sqrt{U_0\left(Z_{R,x}^{-2} +Z_{R,y}^{-2} + 2 k^2\right)/m}$
and the recoil energy $E_\mrm{R}=\hbar^2 k^2/(2m)$.

Furthermore, we used the paraxial approximation $Z_{R,x/y} \gg 1/k$ in the second line, which is
also used to derive the potential for a Gaussian laser beam and, subsequently, an OCB. Hence, the
paraxial approximation is always fulfilled for these types of potentials. Furthermore, we are concerned
with the regime in which our lattice depth is at least several $E_\mrm{R}$, resulting in $\tilde{U}>1$
and thus $Z_{R,x/y}\: l_z^{-1}\gg 1$.

Putting everything together shows that the regime of the numerator of inequality \eqref{eqDenominatorApprox1}
is of the same order of magnitude as that of the last equality in \eqref{eqApproxDenominator}, resulting in
\begin{equation}
\begin{split}
\frac{\sqrt{2} \, \mn{Z}_0(t) \, l_z}{ Z_{R,x}} \lessapprox \frac{q_{c,z}(t)}{Z_{R,x}} +\frac{\frac{1}{\sqrt{2}} 
\tilde{U}^{7/4} l_z}{ Z_{R,x}}
\approx \frac{q_{c,z}(t)}{Z_{R,x}}
\end{split}
\end{equation}

Now, scales on which the approximated denominator and the exponential function change significantly can be compared.
Using inequality \eqref{eqApproxDenominator}, we conclude that the influence of changes in the denominator is
negligible small on the scales on which the exponential functions drops significantly, resulting in the central
approximation
\begin{equation}
\label{eqDenominator}
\sqrt{1+\left[ \frac{\sqrt{2} \, \mn{Z}_0(t) \, l_z}{ Z_{R,x/y}} \right]^2} \approx 1\:.
\end{equation}

Using once again Fa\'{a} di Bruno's representation for Hermite polynomials [cf. Eq.~\eqref{eqHermiteFaadibruno}]
and this last approximation, together with Euler's formula for the cosine function and integral relation
\eqref{eqIntegralRelation}, Eq.~\eqref{eqIntegrationZ1} adopts the approximated form:
\begin{equation}\label{eqIntegrationZ1ApproxFinal}
\begin{split}
\mrm{I}_{z,1}^{\mathbf{n},ap.}(t)
&
=\frac{U_0}{4 \pi} \, \mrm{I}_{r,1}^{\mathbf{n}_r}(0)
\sum_{k_1+2k_2=n_z} (-1)^{n_z + k_1 + k_2}\frac{n_z!}{k_1! k_2!}
2^{k_1}\sum_{l=0}^{k_1}
\begin{pmatrix}
k_1\\
l
\end{pmatrix}
\left[- \frac{ q_{c,z}(t) }{ \sqrt{2} \,  l_z } \right]^{k_1-l} \, D(l).
\end{split}
\end{equation}
For the sake of readability, we introduced the auxiliary function
\begin{equation}\label{eqIntegrationZ1ApproxFinal}
\begin{split}
D(l)& =
\sum_{\lambda = 0}^{\lfloor l/2 \rfloor}
\begin{pmatrix}
l \\
2 \lambda
\end{pmatrix}
\Gamma \left( \lambda + 1/2 \right)
\Bigg[2 \left[ \frac{q_{c,z}(t)}{ \sqrt{2} \, l_z} \right]^{l-2 \lambda}
\\
&
\hspace{0.5cm} + \mrm{exp}\left(-2 \, k^2\, l_z^2\right) \, \mrm{exp} \Big( +2
\mrm{i} k \left[ q_{c,z}(t)-q_0(t) \right] \Big) \left[\frac{q_{c,z}(t)}
{\sqrt{2} \, l_z} + \mrm{i} \, \sqrt{2} \, k \, l_z \right]^{l-2 \lambda}
\\
&
\hspace{0.5cm} + \mrm{exp} \left(-2 \, k^2 \, l_z^2\right) \, \mrm{exp} \Big( -2 \mrm{i}
k [q_{c,z}(t)-q_0(t)] \Big) \left[\frac{q_{c,z}(t)}{ \sqrt{2} \, l_z } - \mrm{i}\, \sqrt{2} k \, l_z \right]^{l-2 \lambda}\Bigg].
\end{split}
\end{equation}

Thus, we have reduced the calculation of $G_\mathbf{n}$ to the numerical evaluation a one-dimensional
integral in the time domain. The relative difference between the results of the full numerical integration
and our approximated solutions was verified to be of the order of $10^{-5}$. At the same time, our
approximate numerical integration is around $15$ times faster than obtaning the numerically-exact solution.

\section{Derivation of the expression for $\mathbf{K}_n$} \label{derivationK_n}
Here we are concerned with the calculation of $\mathbf{K}_{\mathbf{n}}$. To begin with, the gradient of
$\mrm{H_S}$ with respect to $\lambda$ was computed. Use has also been made of the fact that
the substitution $q_0(\boldsymbol{\lambda};t) =  q_0(\boldsymbol{\lambda}_0;t) +  f(\boldsymbol{\alpha};t)$
implies that
\begin{equation}
\begin{split}
\boldsymbol{\nabla}_\lambda q_0(\boldsymbol{\lambda};t) =
\boldsymbol{\nabla}_\alpha f(\boldsymbol{\alpha}; t) \:.
\end{split}
\end{equation}
By inserting the transport modes of the 3D harmonic oscillator, written in the coordinate representation,
into Eq.~\eqref{eqExpressionK}, we obtain
\begin{equation}\label{eqInitialKVec}
\begin{split}
\mathbf{K}_{\mathbf{n}}
&
= \int_0^{t_f} \, dt \, \int_{-\infty}^\infty \, dZ \, \int_{-\infty}^\infty \, dY \, \int_{-\infty}^\infty \, dX
\frac{  -\boldsymbol{\nabla}_\alpha f(\boldsymbol{\alpha}; t) }{\sqrt{2^n n_x! n_y! n_z! \pi^3}} \mrm{exp}
\left[ \mathrm{i} \left( n_x \omega_x + n_y \omega_y + n_z \omega_\mrm{z} \right) t \right]
\\
&
\hspace{0.5cm} \times \frac{C \, P_0}{w_x \left[ Z_0(t) \right]w_y \left[ Z_0(t) \right] }  \mrm{H}_{n_x} \left(
X\right) \mrm{H}_{n_y} \left( Y \right) \mrm{H}_{n_z} \left[ Z_C(t) \right] \, \mrm{exp} \left[ -X^2  - Y^2  -
Z_C(t)^2 \right]
\\
&
\hspace{0.5cm} \times \, \mrm{exp} \left[ -4 \left(\frac{ X^2 l_\mrm{x}^2}{ w_x[Z_0(t)]^2} + \frac{ Y^2 \, l_\mrm{y}^2}
{w_y[Z_0(t)]^2} \right) \right]
\Bigg( k \, \mrm{ sin } \left[ 2^{3/2} \, k \, l_z Z_0(t) \right] + \frac{ Z_0(t) }{ 2^{5/2} \, l_z^3 } \, \Big( \mrm{cos} \left[ 2^{3/2} \, k \,
l_z Z_0(t)\right] + 1 \Big)
\\
&
\hspace{0.5cm} \times \Bigg[  \Bigg( 1 - \frac{8 X^2\, l_\mrm{x}^2}{  w_x \left[ Z_0(t) \right]^2} \Bigg)
\left( \frac{Z_{R,x}}{ \frac{ Z_{R,x}^2 }{ 2 \, l_z^2 } + Z_0(t)^2} \right)^2
+ \Bigg( 1 - \frac{ 8 Y^2 \, l_\mrm{y}^2}{ w_y \left[ Z_0(t) \right]^2} \Bigg) \left( \frac{Z_{R,y}}{
\frac{Z_{R,y}^2}{2 \, l_z^2} + Z_0(t)^2} \right)^2 \Bigg] \Bigg) \:.
\end{split}
\end{equation}

\subsection{Integration over the transverse directions} \label{SubSubSecKIntegrartionXY}
By analogy to what was done in Sec.~\ref{SubSubSecIntegrationXY}, we first treat the integration
in $X$ and $Y$. In other words, we are considering the integral
\begin{equation} \label{eqKIntegrationR}
\begin{split}
\mrm{\tilde{I}}_r^{\mathbf{n}_r}\left[Z_0(t)\right]
&
= \int_{-\infty}^\infty \, dY \,\int_{-\infty}^\infty \, dX \, \mrm{exp} \left( - X^2 - Y^2\right)
\mrm{exp} \left[ -4 \left( \frac{ X^2 \, l_\mrm{x}^2}{ w_x \left[ Z_0(t) \right]^2}  +  \frac{  Y^2 \,
l_\mrm{y}^2}{w_y \left[ Z_0(t) \right]^2} \right)\right]
\\
&
\hspace{0.5cm} \times \mrm{H}_{n_x} \left( X \right) \mrm{H}_{n_y} \left( Y \right) \Bigg( k \,
\mrm{ sin } \left[ 2^{3/2} \, k \, l_z Z_0(t) \right] + \frac{ Z_0(t) }{ 2^{5/2} \, l_z^3 } \, 
\left[ \mrm{cos} \left( 2^{3/2} \, k \, l_z
Z_0(t)\right) + 1 \right]
\\
&
\hspace{0.5cm} \times \Bigg[ \left( \frac{Z_{R,x}}{ \frac{Z_{R,x}^2}{2 \, l_z^2} + Z_0(t)^2} \right)^2
\Bigg( 1 - \frac{8 X^2\, l_\mrm{x}^2}{ w_x \left[ Z_0(t) \right]^2} \Bigg)
+ \left( \frac{Z_{R,y}}{ \frac{Z_{R,y}^2}{2 \, l_z^2} + Z_0(t)^2} \right)^2 \Bigg( 1 - \frac{ 8 Y^2 \,
l_\mrm{y}^2}{  w_y \left[ Z_0(t) \right]^2 } \Bigg) \Bigg] \Bigg).
\end{split}
\end{equation}
The integrations in $X$ and $Y$ are conceptually the same and somewhat similar to what was done in
previous sections. As a consequence, the first term can readily be obtained using previous results:
\begin{equation}\label{eqKIntegrationR1}
\mrm{\tilde{I}}_{r,1}^{\mathbf{n}_r}\left[ Z_0(t) \right]= k \, \mrm{ sin }
\left[ 2^{3/2} \, k\, l_z Z_0(t) \right] \mrm{I}_{r,1}^{\mathbf{n}_r} \left[ Z_0(t) \right]\:.
\end{equation}

The second integral is given by
\begin{equation}\label{eqKIntegrationR2x1Final}
\begin{split}
\mrm{\tilde{I}}_{r,2,x}^{n_x}\left[Z_0(t)\right]
&
= \int_{-\infty}^\infty \, dX \, X^2 \, \mrm{H}_{n_x} \left( X \right) \mrm{exp} \left( -X^2 \right)
\, \mrm{exp} \left( -4 \frac{X^2\, l_\mrm{x}^2}{w_x \left[ Z_0(t) \right]^2} \right)
\\
&
= \sum_{ \substack{k_1 + 2 k_2 = n_x \\ k_1 \mrm{even} } } \frac{n_x!}{k_1! k_2!} (-1)^{k_2} \,
2^{k_1} \, \left( \frac{ w_x \left[ Z_0(t) \right]^2 }{4 \, l_\mrm{x}^2 + w_x \left[ Z_0(t) \right]^2}
\right)^{\frac{k_1 + 3}{2}}
\Gamma \left( \frac{k_1+3}{2} \right).
\end{split}
\end{equation}
where we made use of integral relation \eqref{eqIntegralRelation} and Fa\'{a} di Bruno's representation of
Hermite polynomials \eqref{eqHermiteFaadibruno}. It should be stressed that Eq.~\eqref{eqKIntegrationR2x1Final}
is equal to zero for odd $n_x$ due to the symmetry of the integral, akin to the $X$-integration for $G_{\mathbf{n}}$
[cf. Eq.~\eqref{eqGeneralGRIntegration}]. The $Y$-integration entails similar steps. Putting it all together,
the full form of the second term of Eq.~\eqref{eqInitialKVec} reads
\begin{equation}
\begin{split}
\label{eqKIntegrationR2x1Final}
\mrm{\tilde{I}}_{r,2}^{\mathbf{n}_r} \left[ Z_0(t) \right] =
&
-\sum_{i\in\{\mrm{x,y}\}} C_i\left[ Z_0(t) \right] \frac{ 2^{3/2}\, Z_0(t)\, l_z \, l_i^2}{ w_i\left[ Z_0(t) \right]^2
\sqrt{\pi}} \, \mrm{I}_{r,1}^{0,n_i}\left[ Z_0(t) \right]
\\
&
\times \sum_{ \substack{k_1 + 2 k_2 = n_i \\ k_1 \mrm{even} } } (-1)^{k_2} \frac{ n_x! }{ k_1! k_2! } 2^{k_1+2}
\Gamma \left( \frac{ k_1 + 3 }{2} \right) \, \left( \frac{  w_i\left[ Z_0(t) \right]^2 }{4 \, l_i^2 +  w_i \left[
Z_0(t) \right]^2} \right)^{ \frac{k_1}{2} + 1 }
\\
&
+ \Big( C_x \left[ Z_0(t) \right] + C_y\left[ Z_0(t) \right] \Big)
\, \mrm{I}_{r,1}^{\mathbf{n}_r}\left[ Z_0(t) \right]\:,
\end{split}
\end{equation}
with the factors
\begin{equation}
C_i(Z) = \frac{1}{8 \, l_z^{4}} \left[ \mrm{cos} \left( 2^{3/2} \, k\, l_z Z \right) + 1 \right]
\left( \frac{ Z_{R,i} }{ \frac{ Z_{R,i}^2 }{ 2 \, l_z^2 }+ Z^2 } \right)^2 \:.
\end{equation}
It should be stressed that the expressions for $\mrm{\tilde{I}}_{r,1}^{\mathbf{n}_r}$ and
$\mrm{\tilde{I}}_{r,2}^{\mathbf{n}_r}$ are only valid for even values of $n_x$ and $n_y$,
otherwise they are equal to zero.

\subsection{Integration over the longitudinal direction}\label{SubSubSecKIntegrationZ}
The most general form we can obtain without approximation of the integrand for $\mathbf{K}_\mathbf{n}$
is given by
\begin{equation}\label{eqKIntegrationIntermediate}
\begin{split}
\mathbf{K}_{\mathbf{n}}
&
= - \, \int_0^\mrm{t_f} dt \, \int_{-\infty}^\infty \, dZ \, \frac{U_0 }{\sqrt{2^n n_x! n_y! n_z! \pi^3}} \,
\mrm{exp} \left[ \mrm{i} \left( n_x \omega_x + n_y \omega_y + n_z \omega_\mrm{z} \right) t \right]
\boldsymbol{\nabla}_\alpha f(\boldsymbol{\alpha}; t)
\\
&
\hspace{0.5cm} \times \frac{ \mrm{H}_{n_z} \left[ Z_C(t) \right]  \mrm{exp} \left[ - Z_C(t)^2 \right] }
{\sqrt{1 + \left[ \frac{\sqrt{2} \, Z_0(t)\, l_z}{ Z_{R,x}} \right]^2} \sqrt{1+\left[ \frac{\sqrt{2} \, Z_0(t)\, l_z}{ Z_{R,y}}
\right]^2}}
\Bigg( k \, \mrm{ sin } \left[ 2^{3/2} \, k \, l_z Z_0(t) \right] \, \mrm{I}_{r,1}^{\mathbf{n}_r} \left[ Z_0(t)
\right] + \mrm{\tilde{I}}_{r,2}^{\mathbf{n}_r} \left[ Z_0(t) \right]  \Bigg).
\end{split}
\end{equation}
Hence, we need to evaluate the following integral in the $Z$ direction:
\begin{equation}\label{eqKIntegrationZ}
\begin{split}
\mrm{\tilde{I}}_z^\mathbf{n}(t)
&
= - \int_{-\infty}^\infty \, \frac{d Z \,  \mrm{H}_{n_z} \left[ Z_C(t)  \right]  \mrm{exp} \left[ - Z_C(t)^2
\right]}{\sqrt{1+\left[ \frac{\sqrt{2} \, Z_0(t) \, l_z}{ Z_{R,x}} \right]^2} \sqrt{1+\left[ \frac{\sqrt{2} \, Z_0(t) \, l_z}{ Z_{R,y}}
\right]^2}}
\Bigg( k \, \mrm{ sin } \left[ 2^{3/2} \, k \, l_z Z_0(t) \right] \, \mrm{I}_{r,1}^{\mathbf{n}_r} \left[ Z_0(t) \right]
+ \mrm{\tilde{I}}_{r,2}^{\mathbf{n}_r} \left[ Z_0(t)\right] \Bigg) \:.
\end{split}
\end{equation}
While this integral cannot be computed analytically, we are able to approximate it.

The same procedure to obtain an approximate solution for the $Z$ integral in the first auxiliary function
$G_\mathbf{n}$ can be used to find an approximate result for the integration of Eq.~\eqref{eqKIntegrationZ}.
Repeating the same steps -- that is, approximation of the $z$-dependent denominator -- using the Euler formula
and the Fa\'{a} di Bruno representation for Hermite polynomials [cf. Eq.~\eqref{eqHermiteFaadibruno}], we
obtain the final form for the first term
\begin{equation}\label{eqIntegrationTildeZ1ApproxFinal}
\begin{split}
\mrm{\tilde{I}}_{z,1}^{\mathbf{n},ap.}(t)
&
= \frac{ \mrm{i} k }{2} \, \mrm{I}_{r,1}^{\mathbf{n}_r}(0) \, \mrm{exp}\left( -2 \, k^2\, l_z^2 \right)
\sum_{k_1+2k_2=n_z}\frac{ 2^{k_1} n_z!}{k_1! k_2!} (-1)^{n_z + k_1 + k_2 + 1}
\sum_{l=0}^{k_1}
\begin{pmatrix}
k_1\\
l
\end{pmatrix}
\\
&
\hspace{0.5cm} \times
\left[- \frac{ q_{c,z}(t) }{ \sqrt{2} \,  l_z }\right]^{k_1-l}
\sum_{\lambda}^{\lfloor l/2 \rfloor}
\Bigg[ \mrm{exp} \Big( 2 \mrm{i} k \left[ q_{c,z}(t)-q_0(t) \right] \Big)  \left[ \frac{ q_{c,z}(t) }{ \sqrt{2} \, l_z }
+ \mrm{i} \sqrt{2} \, k \, l_z \right]^{l-2 \lambda}
\\
&
\hspace{0.5cm} - \mrm{exp} \Big( -2 \mrm{i} k \left[ q_{c,z}(t) - q_0(t) \right] \Big)
\left[\frac{ q_{c,z}(t) }{ \sqrt{2} \, l_z }
- \mrm{i} \sqrt{2} \, k \, l_z\right]^{l-2 \lambda}\Bigg]
\begin{pmatrix}
l \\
2 \lambda
\end{pmatrix}
\Gamma \left( \lambda + 1/2 \right)
\end{split}
\end{equation}
and the second term of Eq.~\ref{eqKIntegrationZ}.
\begin{equation}
\begin{split}\label{eqKIntegrationZ2ApproxFinal}
\mrm{\tilde{I}}_{z,2}^{\mathbf{n},ap.}(t) & =
-\sum_{i\in\{\mrm{x,y}\}}
\frac{2 \, l_i^2}{Z_{R,i}^2  w_{i,0}^2 \sqrt{\pi}} \, \mrm{I}_{r,1}^{0,n_i}(0)\sum_{
\substack{k_1 + 2 k_2 = n_i \\ k_1 \mrm{even} } } (-1)^{k_2} \frac{ n_i! }{ k_1! k_2! }
2^{k_1+2} \Gamma \left( \frac{ k_1 + 3 }{2} \right)
\\
&
\hspace{0.5cm} \times
\left( \frac{ w_{i,0}^2 }{4 \, l_i^2 +  w_{i,0}^2} \right)^{ \frac{k_1}{2} + 1 }
\Bigg(\sum_{\tilde{k}_1+2\tilde{k}_2=n_z} \frac{n_z!}{\tilde{k}_1! \tilde{k}_2!} (-1)^{n_z
+ \tilde{k}_1 + \tilde{k}_2} 2^{\tilde{k}_1 - 1}
\sum_{l=0}^{\tilde{k}_1} \left[- \frac{ q_{c,z}(t) }{ \sqrt{2} \,  l_z } \right]^{\tilde{k}_1-l}
\\
&
\hspace{0.5cm} \times
\begin{pmatrix}
\tilde{k}_1\\
l
\end{pmatrix}
 \left[D(l+1)- \frac{ q_0(t) }{ \sqrt{2} \, l_z } \, D(l)\right]\Bigg)
+ \left( \frac{1}{Z_{R,x}^2} + \frac{1}{Z_{R,y}^2} \right) \, \mrm{I}_{z,1}^{\mathbf{n},ap.}(0)
\, \mrm{I}_{r,1}^{\mathbf{n}_r}(0) \:.
\end{split}
\end{equation}
The computational speedup and the accuracy of the final result are of the same order as in the
aforementioned approximation for $G_\mathbf{n}$ [cf. Sec.~\ref{SubSubSecApproxGn}].

\section{Parameters of the auxiliary function $f(\alpha;t)$}\label{tableAppendix}
The values of the parameters $\tilde{a}_{n,k}$ in Eq.~\eqref{eqSolutionvectorF},
rounded to the accuracy of $10^{-8}$, are listed in Table I.

\begin{table}[H]
\centering
\begin{tabular}{c|cccccc}
\diagbox{$n$}{$k$} & 1 & 2 & 3 & 4 & 5 & 6 \\
\hline
3  & \num{3268.0278}		& \num{-1764.7350}		& \num{1361.6782}		& \num{-1021.2587}		& \num{705.89400}			& \num{-544.67130}
\\
4  & \num{-42974.565}	& \num{29382.838} 		& \num{-24260.567}		& \num{18791.160}		& \num{-13235.513} 		& \num{10339.677}
\\
5  & \num{238311.85}	& \num{-188292.32}		& \num{168031.09} 		& \num{-135594.78} 	& \num{97923.184}		& \num{-77792.678}
\\
6  & \num{-731080.51}	& \num{636579.13}		& \num{-607620.55} 	& \num{512478.95}		& \num{-381148.45}		& \num{309119.12}
\\
7  & \num{1362055.0}	& \num{-1270967.0}	& \num{1282059.8} 	& \num{-1128042.6}	& \num{865989.78}		& \num{-719297.45}
\\
8  & \num{-1583096.2}	& \num{1555055.1}		& \num{-1640810.9}	& \num{1500138.2}		& \num{-1188990.2}	& \num{1013733.1}
\\
9  & \num{1124047.2}	& \num{-1148396.4} 	& \num{1257158.1} 	& \num{-1189045.8}	& \num{971849.37}		& \num{-851598.11}
\\
10 & \num{-446816.56}	& \num{470792.08}		& \num{-531275.79}		& \num{517653.33} 		& \num{-435482.68}		& \num{392326.73}
\\
11 & \num{76285.754}	& \num{-82388.614}		& \num{95357.192} 		& \num{-95357.192}		& \num{82388.614}		& \num{-76285.754}
\end{tabular}
\caption{Approximated parameters $\tilde{a}_{n,k}$ for Eq.~\eqref{eqSolutionvectorF}.
The relative differences between approximated and exact values are of the order of \num{e-8}.
For $n<3$ the parameter $\tilde{a}_{n,k}$ is equal to zero.}
\label{tabFParameters}
\end{table}

\twocolumngrid

\end{document}